\documentclass[preprintnumbers,showpacs,amsmath,amssymb,floatfix,prd,onecolumn,superscriptaddress,nofootinbib]{revtex4}
\usepackage{bbm}
\usepackage{graphicx}
\usepackage{epsfig}
\usepackage{bm}
\usepackage{amsfonts}
\usepackage{epstopdf}
\usepackage{multirow}

\begin{document}

\title{Non-Gaussianity with Lagrange Multiplier Field in the Curvaton Scenario}

\author{Chao-Jun Feng}
\email{fengcj@shnu.edu.cn} \affiliation{Shanghai United Center for Astrophysics (SUCA), \\ Shanghai Normal University,
    100 Guilin Road, Shanghai 200234, P.R.China}

\author{Xin-Zhou Li}
\email{kychz@shnu.edu.cn} \affiliation{Shanghai United Center for Astrophysics (SUCA),  \\ Shanghai Normal University,
    100 Guilin Road, Shanghai 200234, P.R.China}

\begin{abstract}
 In this paper, we will use $\delta \mathcal{N}$-formalism to calculate the primordial curvature perturbation for the curvaton model with a Lagrange multiplier field.  We calculate the non-linearity parameters $f_{NL}$ and $g_{NL}$ in  the sudden-decay approximation in this kind of model,  and we find that one could get a large non-Gaussinity even if the curvaton dominates the total energy density before it decays, and this property will make the curvaton model much richer. We also calculate the probability density function of the primordial curvature perturbation in the sudden-decay approximation, as well as some moments of it. 
 \end{abstract}


 \maketitle

\section{Introduction}\label{sec:intro}
Inflation has been remarkably successful in explaining the properties of the universe and the origin of the primordial
perturbation \cite{inflation}, which is thought of as the seed of the large scale structures. So far, there still a lot of discussions and works on inflation, such as \cite{fengli}. A single-field inflation predicts a nearly Gaussian distribution of the primordial power spectrum \cite{Maldacena:2002vr}. On the other hand, multi-field models of inflation can lead to a large deviation from the Gaussian distribution, which may be observed in the future observations \cite{dwang}. In fact, the multi-field models generate the non-Gaussianity due to the non-trivial classical dynamics on superhorizon scales.  Since the gravitational dynamics could introduce significant non-linearities that would contribute to the final non-Gaussianity in the large scale of CMB anisotropies, the CMB non-Gaussianity opens a window to probe the physics of the early universe. 

There are many mechanisms to generate a large local-type non-Gaussinities,  and one of them is the curvaton scenario \cite{curvaton}. In this kind of model, there would be another, weakly coupled, light inhomogeneous scalar field  called curvaton, whose energy density could be neglected during inflation, while the early Universe is dominated by inflaton. After the end of inflation, the energy of inflaton converted  into radiations and the Hubble parameters decreases. During the evolution of the curvaton, its energy density goes like $\propto a^{-3}$, which increases with respect to that of radiations  $\propto a^{-4}$. Therefore, the curvaton can dominate the energy density of the Universe later. When the Hubble parameter becomes the same order of the curvation decay rate, the energy of the curvaton would be converted into radiations. Finally, the curvation is supposed to completely decay into thermalized radiations before primordial nucleosynthesis, in the meanwhile, the perturbations of the curvaton  become the final adiabatic curvature perturbations that seed the matter and radiation density fluctuations observed in the Universe. This kind of non-Gaussianity can be described by some non-linearity parameters $f_{NL}$, $g_{NL}$, etc. defined below.  For recent progress on the curvaton model, see Ref.~\cite{qghuang}\cite{piao}\cite{other}.

At first, we expand the curvature perturbation as
\begin{equation}
	\zeta(t, \mathbf{x}) = \zeta_{1}(t, \mathbf{x})  + \sum_{n=2}^{\infty} \frac{1}{n!} \zeta_{n}(t, \mathbf{x}) \,,
\end{equation}
where the probability density function (pdf) of the first order term $\zeta_{1}$ is Gaussian, while the higher order terms give rise to a non-Gasussian pdf of the full $\zeta$. As usual, the non-linearity parameters $f_{NL}$ and $g_{NL}$ are defined by
\begin{equation}
	\zeta = \zeta_{1} + \frac{3}{5}f_{NL} \zeta_{1}^{2} + \frac{9}{25} g_{NL}\zeta_{1}^{3} + \mathcal{O}(\zeta_{1}^{4}) \,,
\end{equation}
or, equivlently
\begin{equation}\label{eqn:fnlgnl}
	f_{NL} = \frac{ 5\zeta_{2} }{ 6\zeta_{1}^{2} } \,, \quad g_{NL} = \frac{ 25 \zeta_{3} }{ 54\zeta_{1}^{3} } \,,
\end{equation}
where the numerical factors arise in order to be consistent with the Bardeen potential on large scales.  The upper bound from the WMAP 3yr data \cite{WMAP3} is $|f_{NL}|<114$ , the bound from WMAP 5yr data \cite{WMAP5} is $-9<f_{NL}<111$ at $2\sigma$ level and the constraint from WMAP 7yr data \cite{WMAP7} is $f_{NL}=32\pm21$ at $1\sigma$ level. The correlation functions of Fourier transformation of $\zeta$ are used to define the primordial power spectrum, bispectrum and trispectrum as 
\begin{eqnarray}
	\langle \zeta(\mathbf{k_{1}})\zeta(\mathbf{k_{2}})\rangle 
		&=& (2\pi)^{3} \mathcal{P}_{\zeta}(k_{1}) \delta^{3}\bigg( \sum_{n=1}^{2}\mathbf{k}_{n}\bigg) \,, \\
	\langle \zeta(\mathbf{k_{1}})\zeta(\mathbf{k_{2}})\zeta(\mathbf{k_{3}})\rangle 
		&=& (2\pi)^{3} \mathcal{B}_{\zeta}(\mathbf{k}_{1}, \mathbf{k}_{2}) \delta^{3}\bigg( \sum_{n=1}^{3}\mathbf{k}_{n}\bigg) \,, \\
	\langle \zeta(\mathbf{k_{1}})\zeta(\mathbf{k_{2}})\zeta(\mathbf{k_{3}})\zeta(\mathbf{k_{4}})\rangle &=& 
		(2\pi)^{3} \mathcal{T}_{\zeta}(\mathbf{k}_{1}, \mathbf{k}_{2},\mathbf{k}_{3} )\delta^{3}\bigg( \sum_{n=1}^{4}\mathbf{k}_{n}\bigg) \,.
\end{eqnarray}
Thus, we have
\begin{eqnarray}
	  \mathcal{B}_{\zeta}(\mathbf{k}_{1}, \mathbf{k}_{2})  &=& \frac{6}{5}f_{NL}\bigg[\mathcal{P}_{\zeta}(k_{1})\mathcal{P}_{\zeta}(k_{2}) + 2\text{perms}\bigg] \,,\\
	  \nonumber
	  \mathcal{T}_{\zeta}(\mathbf{k}_{1}, \mathbf{k}_{2},\mathbf{k}_{3} ) &=&
	  \frac{18}{25} f_{NL}^{2}\bigg[\mathcal{P}_{\zeta}(k_{1})\mathcal{P}_{\zeta}(k_{2})\mathcal{P}_{\zeta}(|\mathbf{k}_{1}-\mathbf{k}_{2}|) + 23\text{perms}\bigg] \\ 
	  &+&  \frac{54}{25}g_{NL}\bigg[\mathcal{P}_{\zeta}(k_{1})\mathcal{P}_{\zeta}(k_{2})\mathcal{P}_{\zeta}(k_{3}) + 3\text{perms}\bigg]  \,.
\end{eqnarray} 

In this paper, these non-linearity parameters will be calculated  in the curvaton scenario with a Lagrange multiplier field,  which is described by the following action \cite{lagrange}
\begin{equation}\label{equ:action}
	S = \int d^{4}x \sqrt{-g} \bigg[ K(\varphi, X) + \lambda \bigg(X - V(\varphi) \bigg) \bigg] \,,
\end{equation}
where the field $\lambda$ is a ``Lagrange multiplier'' without a kinetic term and the scalar field $\varphi$ could be a curvaton. Here, 
\begin{equation}
	X \equiv \frac{1}{2}g^{\mu\nu}\nabla_{\mu}\varphi\nabla_{\nu}\varphi \,,
\end{equation}
is a standard kinetic term for the field $\varphi$, $K(\varphi, X)$ is arbitrary function of $\varphi$ and $X$, and $V(\varphi)$ is an arbitrary function of the scalar field $\varphi$. The equations of motion for $\lambda$ and $\varphi$ are given by
\begin{eqnarray}
	X - V(\varphi) &=& 0  \,, \label{equ:EOM1.1}\\
	K_{\varphi}-\nabla_{\mu}(K_{X}\nabla^{\mu}\varphi)- \lambda V_{\varphi} - \nabla_{\mu}(\lambda\nabla^{\mu}\varphi)&=&0 \,. \label{equ:EOM1.2}
\end{eqnarray}
Here and after we denote the partial derivatives by subscripts.  And the energy-momentum tensor is
\begin{equation}\label{equ:EMT}
	T_{\mu\nu} = (K_{X} + \lambda)\nabla_{\mu}\varphi \nabla_{\nu}\varphi - Kg_{\mu\nu}\,.
\end{equation}

In the spatially flat FRW universe with the metric $ds^{2}=dt^{2}-a^{2}(t)dx^{2}$,  we consider the homogeneous Lagrange multiplier field so that  Eqs. (\ref{equ:EOM1.1}) and  (\ref{equ:EOM1.2}) reduce to
\begin{eqnarray}
	 \dot\varphi^{2} -  2V(\varphi) &=&0  \,, \label{equ:EOM2.1}\\
	\frac{K_{\varphi}}{\sqrt{2V}} - \frac{V_{\varphi}}{\sqrt{2V}} \bigg(2VK_{XX}+K_{X}+2\lambda \bigg) - \sqrt{2V}K_{X\varphi} -3H(K_{X}+\lambda)- \dot\lambda      &=& 0 \,. \label{equ:EOM2.2}
\end{eqnarray}
And also, the energy density and pressure are given by
\begin{eqnarray}
	\rho(\lambda, \varphi) &=& 2V(\varphi) (K_{X} + \lambda)-K \,, \\
	p(\varphi) &=& K\big(\varphi, V(\varphi)\big) \,.
\end{eqnarray}

In this paper, we will consider two interesting cases that studied in \cite{lagrange}. The first case is $K_{\varphi}=0$ and $V= \text{const.}$, in which case, the energy-momentum corresponds to a mixture of a cosmological constant and pressureless dust, so we call this model the $L-\lambda\varphi$ model. 
From Eqs. (\ref{equ:EOM2.1}) and  (\ref{equ:EOM2.2}), we get
\begin{equation}
	\dot\varphi = \sqrt{2V}\,, \quad \dot\lambda+3H(K_{X}+\lambda)= 0\,,
\end{equation}
with solution
\begin{equation}
	\varphi = \sqrt{2V} t \,, \quad  \lambda = \frac{\rho_{0}}{2V a^{3}} - K_{X},,
\end{equation}
where $\rho_{0}$ is an integration constant. Thus,  the energy density and pressure  are given by
\begin{equation}
	\rho = \frac{\rho_{0}}{a^{3}} - p \,, \quad p = \text{const} \,.
\end{equation}
The other case is  $K=0$ with arbitrary $V$,  and in this model the energy density evolves exactly the same as dust or matter, so we call this model $D-\lambda\varphi$ model.  The equations of motion for this model are 
\begin{equation}
	\dot\varphi = \sqrt{2V(\varphi)}\,, \quad \ddot\varphi + 3H\dot\varphi + V_{\varphi} + \lambda^{-1}\dot\lambda\dot\varphi= 0\,,
\end{equation}
and by using the relation $\ddot\varphi = 2V(\varphi)$, we get the solution 
\begin{equation}
	\lambda = \frac{\rho_{0}}{a^{3} \dot\varphi^{2}} = \frac{\rho_{0}}{2V(\varphi)a^{3}} \,.
\end{equation}
Therefore the energy density and pressure  are given by
\begin{equation}
	\rho = \frac{\rho_{0}}{a^{3}}  \,, \quad p =0\,.
\end{equation}
which is exactly the behavior of the pressureless dust or the cold dark matter. It should be noticed that the expression of the energy density  is independent of the explicit form of the scalar potential.

It should be noticed that in a single-field inflation, the prediction of the non-linearity parameter is related to the tilt of the power spectrum \cite{Maldacena:2002vr}, so if a large local-type non-Gaussianity is confirmed by the future cosmological observations and high level data analysis, it strongly implies that the physics of the early Universe is more complicated than the simple single-field slow-roll inflation. In this paper, we will use $\delta \mathcal{N}$-formalism \cite{deltan} to calculate the primordial curvature perturbation for the curvaton model with a Lagrange multiplier field. And we find that, in this kind of model, one could get a large non-Gaussinity even if the curvaton dominates the total energy density before it decays. Our paper is organized as follows. In Sec.~\ref{sec:NonPer}, we calculate the non-linearity parameters for the curvaton itself and in Sec.~\ref{sec:SDNC}, we derive in the sudden-decay approximation a non-linear equation that relates the primordial curvature perturbation $\zeta$ to the curvaton curvature perturbation $\zeta_{\varphi}$. Solving this equation order by order, we obtain the non-linearity parameters $f_{NL}$ and $g_{NL}$ in  the sudden-decay approximation and we also compare the results with the usual curvaton model \cite{curvaton}. In Sec.~\ref{sec:pdf}, we calculate the pdf and some moments of $\zeta$ in the sudden-decay approximation. In the final section, we will give some conclusions and discussions.

\section{Non-Linear Perturbation with the Lagrange Multiplier Field}\label{sec:NonPer}

The primordial density perturbation can be described in terms of the non-linear curvature perturbation on uniform density hypersurface \cite{Lyth:2004gb}
\begin{equation}\label{uniform density}
	\zeta(t, \mathbf{x}) = \delta N (t, \mathbf{x}) + \frac{1}{3}\int_{\bar\rho(t)} ^{\rho (t, \mathbf{x})} \frac{d\tilde\rho}{\tilde\rho + \tilde p} \,,
\end{equation}
where $N=\int Hdt$ is the amount of local expansion, and $\rho^{i}$, $p^{i}$ are the local energy and pressure respectively. Here, $\bar\rho$ is the homogeneous energy density in the background model, while $\tilde\rho$ and $\tilde p$ is the local density and local pressure. 

\subsection{The $L-\lambda\varphi$ model}
In this model, we will assume that the constant pressure $p$ is much smaller than the energy density, so the scalar field behaves much like the pressureless dust, thus we have the non-linear curvature perturbation on uniform-curvaton density surfaces is given by \cite{Lyth:2004gb}
\begin{equation}
	\zeta(t, \mathbf{x})_{\varphi} = \delta N (t, \mathbf{x}) + \int_{\bar\rho_{\varphi}(t)} ^{\rho_{\varphi} (t, \mathbf{x})} \frac{d\tilde\rho_{\varphi}}{3\tilde\rho_{\varphi} } \,.
\end{equation}
Hence, the density of the scalar field $\varphi$ on spatially-flat hypersurfaces is given by
\begin{equation}\label{equ:flat fluc 1}
	\rho_{\varphi} |_{\delta N = 0} = e^{3\zeta_{\varphi}} \bar\rho_{\varphi} \,.
\end{equation}

Generally, we can expand any field 
\begin{equation}
	\varphi = \bar\varphi + \sum_{n=1}^{\infty} \frac{1}{n!}\delta_{n}\varphi \,,
\end{equation}
where $\bar\varphi$ is the homogeneous background filed. On the other sider, the quantum fluctuations in a weakly coupled field could be well described by a Gaussian random field \cite{Sasaki:2006kq}. So for such fields, one can only keep the first order perturbation $\delta_{1}\varphi $ and the higher order perturbations $\delta_{n}\varphi$ for $n>1$ that describe non-Gassian perturbations of any field could be neglected. However, here we also want to estimate the effect of the non-linear quantum fluctuations in the curvaton field and Lagrange multiplier field at Hubble exit during inflation, we will keep to the third order of the fluctuations as 
\begin{eqnarray}
	\varphi_{*} &=& \bar\varphi_{*} + \delta_{1}\varphi_{*} +  \frac{1}{2}\delta_{2}\varphi_{*}+\frac{1}{6} \delta_{3}\varphi_{*}\, \\
	\lambda_{*} &=& \bar\lambda_{*} + \delta_{1}\lambda_{*} +  \frac{1}{2}\delta_{2}\lambda_{*}+\frac{1}{6} \delta_{3}\lambda_{*}\,
\end{eqnarray}
where $*$ denotes that the quantities are evaluated at the Hubble exit during inflation. Therefore, we get the density fluctuation of the curvaton as
\begin{equation}\label{equ:rho1}
	\rho_{\varphi*} = \bar\rho_{\varphi*} + 2V\left( \delta_{1}\lambda_{*} +  \frac{1}{2}\delta_{2}\lambda_{*}+\frac{1}{6} \delta_{3}\lambda_{*}\right) \,,
\end{equation}
where $\bar\rho_{\varphi*} = 2V(K_{X} + \lambda_{*})-p \approx 2V(K_{X} + \lambda_{*})$. Therefore, order by order,  from Eq. (\ref{equ:flat fluc 1}) we have
\begin{equation}
	\delta_{1}\rho_{\varphi*} = 2V \delta_{1}\lambda{*} \,,\quad 
	\delta_{2}\rho_{\varphi*} = 2V \delta_{2}\lambda{*} \,, \quad
	\delta_{3}\rho_{\varphi*} = 2V \delta_{3}\lambda{*} \,,
\end{equation}
and  we also have $e^{3\zeta_{\varphi}} = \rho_{\varphi*}/\bar\rho_{\varphi*}$, then
\begin{eqnarray}
	\zeta_{\varphi 1} &=& \frac{2V}{3\bar\rho_{\varphi*}} \delta_{1}\lambda_{*} \,, \label{equ:zeta1}\\
	\zeta_{\varphi 2} &=& \frac{2V}{3\bar\rho_{\varphi*}} \delta_{2}\lambda_{*} - 3\zeta_{\varphi 1}^{2}  = 3\left(a_{1}-1\right)\zeta_{\varphi 1}^{2}\,,  \label{equ:zeta2}\\
	\zeta_{\varphi 3} &=& \frac{2V}{3\bar\rho_{\varphi*}} \delta_{3}\lambda_{*} + 9(2-3a_{1})\zeta_{\varphi 1}^{3} = 9(2-3a_{1} + b_{1})\zeta_{\varphi 1}^{3} \,,  \label{equ:zeta3}
\end{eqnarray}
where we have defined
\begin{equation}
	a_{1} = \frac{\bar\rho_{*}}{2V} \frac{ \delta_{2}\lambda_{*} }{( \delta_{1}\lambda_{*})^{2}} \,, \quad b_{1} =  \frac{\bar\rho_{*}^{2}}{4V^{2}} \frac{ \delta_{3}\lambda_{*} }{( \delta_{1}\lambda_{*})^{3}} \,,
\end{equation}
to  estimate the effect of the non-linear quantum fluctuations of the Lagrange multiplier field, and if $a_{1}\sim0$ and $b_{1}\sim0$, this field is almost Gaussian. Using Eqs.~(\ref{equ:zeta1})-(\ref{equ:zeta3}), one can express the non-linearity parameters for the curvaton perturbation analogous to Eq. (\ref{eqn:fnlgnl}) as
\begin{equation}\label{fgNL1}
	f^{\varphi}_{NL} = \frac{5}{2}(a_{1}-1) \,, \quad g^{\varphi}_{NL} = \frac{25}{6}(2-3a_{1}+b_{1}) \,.
\end{equation}
Here we find $f^{\varphi}_{NL}=-5/2$ and $g^{\varphi}_{NL}=25/3$ for a Gaussian $\lambda$ field, while  curvaton field itself could be Gaussian or non-Gaussian, and we also have the relation $10f^{\varphi}_{NL} + 3g^{\varphi}_{NL}=0$. It is worth to note that in this model, the fluctuation of the curvaton field do not contribute to the non-linearity parameters since $V$ is a constant.

\subsection{The $D-\lambda\varphi$ model}
In this model, the density fluctuation is given by
\begin{equation}\label{equ:rho2}
	\rho_{\varphi*} = \bar\rho_{\varphi*} \left(1 +\bar \lambda_{*}^{-1}\delta\lambda_{*}+ \bar V_{*}^{-1}\delta V_{*}\right)\,,
\end{equation}
where $\bar V_{*}=V(\bar \varphi_{*})$ , $\bar\rho_{\varphi*}  = 2\bar V_{*} \bar \lambda_{*}$,
\begin{equation}
	\delta\lambda_{*} =\delta_{1}\lambda_{*} +  \frac{1}{2}\delta_{2}\lambda_{*}+\frac{1}{6} \delta_{3}\lambda_{*} \,,
\end{equation}
and
\begin{eqnarray}
	\nonumber
	\delta V_{*} &=& \bar V'_{*}\delta_{1}\varphi_{*} + \frac{1}{2} \bar V'_{*}\delta_{2}\varphi_{*} + \frac{1}{2} \bar V''_{*}(\delta_{1}\varphi_{*})^{2}  \\
			  &+& \frac{1}{6} \bar V'_{*}\delta_{3}\varphi_{*} + \frac{1}{2} \bar V''_{*}\delta_{1}\varphi_{*}\delta_{2}\varphi_{*} +  \frac{1}{6} \bar V'''_{*}(\delta_{1}\varphi_{*})^{3} \,,
\end{eqnarray}
where the prime denotes the derivative with respect to $\varphi$.  From Eq. (\ref{equ:rho2}), one can see that both the fluctuation of $\lambda$ and $\varphi$ fields contribute to the fluctuation of the density, so  in the following, we will consider two limit case: $\delta V_{*}=0 $  and $\delta\lambda_{*} =0$ to close and open one of them and clearly illustrate the contribution of each field to the non-linearity parameters. In the case of $\delta V_{*} =0$  , we get the the curvature perturbation as
\begin{eqnarray}
	\zeta_{\varphi 1} &=& \frac{ \delta_{1}\lambda_{*}}{3\bar \lambda_{*}} \,, 
	\label{equ:zeta12}\\
	\zeta_{\varphi 2} &=&\frac{ \delta_{2}\lambda_{*}}{3\bar \lambda_{*}}  -3\zeta_{\varphi 1}^{2} = 3(a_{2}-1)\zeta_{\varphi 1}^{2} \,,  
	\label{equ:zeta22}\\
	\zeta_{\varphi 3} &=&\frac{ \delta_{3}\lambda_{*}}{3\bar \lambda_{*}} +  9(2-3a_{2})\zeta_{\varphi 1}^{3}= 9(2-3a_{2} + b_{2})\zeta_{\varphi 1}^{3} \,, 
	\label{equ:zeta32}
\end{eqnarray}
where $a_{2} = \bar \lambda_{*} \delta_{2}\lambda_{*}/( \delta_{1}\lambda_{*})^{2}$ and $b_{2}= \bar \lambda_{*} ^{2}\delta_{3}\lambda_{*}/( \delta_{1}\lambda_{*})^{3}$. Here, one can see that these equations are the same as Eqs.~(\ref{equ:zeta1})-(\ref{equ:zeta3}) except for the definition of $a_{i}$ and $b_{i}$.  While in the case of   $\delta\lambda_{*} =0$, we get
\begin{eqnarray}
	\zeta_{\varphi 1} &=& \frac{\bar V'_{*}}{3\bar V_{*}}\delta_{1}\varphi_{*} \,, 
	\label{equ:zeta13}\\
	\zeta_{\varphi 2} &=& \frac{\bar V'_{*}}{3\bar V_{*}}\delta_{2}\varphi_{*} 
		+ 3\left( \frac{ \bar V''_{*} \bar V_{*} } { \bar V_{*}^{'2} } -1\right) \zeta_{\varphi 1}^{2}
		= 3\left( a_{3} + \frac{ \bar V''_{*} \bar V_{*} } { \bar V_{*}^{'2} } -1\right) \zeta_{\varphi 1}^{2}\,,  
	\label{equ:zeta23}\\
	\nonumber
	\zeta_{\varphi 3} &=& \frac{\bar V'_{*}}{3\bar V_{*}}\delta_{3}\varphi_{*} 
		+ 9\left( 2-3a_{3} -3(1-a_{3})\frac{ \bar V''_{*} \bar V_{*} } { \bar V_{*}^{'2} } + \frac{ \bar V'''_{*} \bar V_{*}^{2} } { \bar V_{*}^{'3} } \right) \zeta_{\varphi 1}^{3} \\
		&=& 9\left( 2-3a_{3} + b_{3}-3(1-a_{3})\frac{ \bar V''_{*} \bar V_{*} } { \bar V_{*}^{'2} } + \frac{ \bar V'''_{*} \bar V_{*}^{2} } { \bar V_{*}^{'3} } \right) \zeta_{\varphi 1}^{3} \,,   
	\label{equ:zeta33}
\end{eqnarray}
where 
\begin{equation}
	a_{3} = \frac{\bar V_{*}}{\bar V'_{*}} \frac{\delta_{2}\varphi_{*}}{(\delta_{1}\varphi_{*})^{2}} \,,
	\quad
	b_{3} = \frac{\bar V_{*}^{2}}{\bar V_{*}^{'2}} \frac{\delta_{3}\varphi_{*}}{(\delta_{1}\varphi_{*})^{3}} \,.
\end{equation}
Then,  using Eqs.~(\ref{equ:zeta13})-(\ref{equ:zeta33}), one can express the non-linearity parameters for the curvaton perturbation analogous to Eq. (\ref{eqn:fnlgnl}) as
\begin{equation}\label{fgNl2}
	f^{\varphi}_{NL} = \frac{5}{2}\left( a_{3} + \frac{ \bar V''_{*} \bar V_{*} } { \bar V_{*}^{'2} } -1\right) \,, \quad 
	g^{\varphi}_{NL} = \frac{25}{6}\left( 2-3a_{3} + b_{3}-3(1-a_{3})\frac{ \bar V''_{*} \bar V_{*} } { \bar V_{*}^{'2} } + \frac{ \bar V'''_{*} \bar V_{*}^{2} } { \bar V_{*}^{'3} } \right) \,.
\end{equation}
Now, let's forget about the $a_{3}$ and $b_{3}$ for a while, and take the function $V$ as the power of $\varphi$, namely, $V\sim\varphi^{n}$, then the non-linearity parameters becomes
\begin{equation}\label{newresult}
	f^{\varphi}_{NL} =  - \frac{5}{2n}  \,, \quad
	g^{\varphi}_{NL} = \frac{25}{3n^{2}} \,,
\end{equation}
which could be larger when $n$ is small. And here,  if we take $V\sim e^{m\varphi}$, these parameters would be vanished. With Eq.~(\ref{newresult}), one can find the following relation
\begin{equation}
       10  f^{\varphi}_{NL}+3n g^{\varphi}_{NL} = 0 \,.
\end{equation}
It should be noticed that, the result Eq.~(\ref{newresult}) is evidently distinguishing with that in Ref.~\cite{Huang:2008zj}, in which the author considered a simple power-law potential of the curvaton, i.e. $V\sim \varphi^{n}$, but the equation of state of the curvaton scaclar depends on the value of $n$ as $w=p/\rho = (n-2)/(n+2)$ in their situation. However, in our model, the pressure is always vanished. In a special case $n=2$, both our result and that  in Ref.~\cite{Huang:2008zj} are coincident with the  result  for the quadratic potential of the curvaton \cite{curvaton}. Of course, one can take the other forms of the function $V$, like $V\sim \cosh (m\varphi)$ to generate large non-Gassuaianity even when the curvaton density dominates the universe, as long as $m\varphi_{*}\ll 1$:
\begin{eqnarray}
	\tilde f^{\varphi}_{NL} &=&  \frac{5}{2\sinh^{2}(m\varphi_{*})} \approx \frac{5}{2(m\varphi_{*})^{2}} \,, \\
	\tilde g^{\varphi}_{NL} &=& -\frac{25}{3\sinh^{2}(m\varphi_{*})} \approx -\frac{25}{3(m\varphi_{*})^{2}}\,,
\end{eqnarray}
and one can also find the relation
\begin{equation}
	10 \tilde f^{\varphi}_{NL}+3\tilde g^{\varphi}_{NL} = 0 \,.
\end{equation}

\section{Sudden-decay Approximation and the Primordial Curvature Perturbation}\label{sec:SDNC}
We assume that the curvaton decays on a uniform-total energy density hypersurface and thus from Eq.~(\ref{uniform density}) the perturbed expansion on this hypersurface is $\delta N = \zeta$, where $\zeta$ is the total curvature perturbations at curvaton decay surface. And on this surface, we have 
\begin{equation}\label{total rho}
	\rho_{r}(t_{d},  \mathbf{x}) + \rho_{\varphi}(t_{d},  \mathbf{x}) = \bar\rho(t_{d}) \,.
\end{equation}
Here, we assume that all the curvaton decay products are relativistic, then $\zeta$ is conserved after the curvaton decay.  The local curvaton and radiation density on this decay surface may be inhomogeneous and they have a conserved curvature perturbations when they do not have interactions with each other \cite{Lyth:2004gb} :
\begin{equation}
	\zeta_{i} = \delta N  + \frac{1}{3}\int_{\bar\rho_{i}} ^{\rho_{i}} \frac{d\tilde\rho_{i}}{\tilde\rho_{i} + \tilde p_{i}} \,.
\end{equation} 
Thus, the curvature perturbation related to radiations ($p_{r} = \rho_{r}/3$) is  
\begin{equation}
	\zeta_{r} = \zeta + \frac{1}{4} \ln \left( \frac{\rho_{r}}{\bar\rho_{r}} \right) \,, \quad \text{or} \quad \rho_{r} = \bar\rho_{r} e^{4(\zeta_{r} - \zeta)} \,,
\end{equation}
and
\begin{equation}
	\zeta_{\varphi} = \zeta + \frac{1}{3} \ln \left( \frac{\rho_{r}}{\bar\rho_{r}} \right) \,, \quad \text{or} \quad \rho_{r} = \bar\rho_{r} e^{3(\zeta_{\varphi} - \zeta)} \,.
\end{equation}
Therefore, from Eq. (\ref{total rho}) we have the following relation
\begin{equation}\label{sudden rel}
	(1-\Omega_{\varphi, d})e^{4(\zeta_{r} - \zeta)} +\Omega_{\varphi, d} e^{3(\zeta_{\varphi} - \zeta)} = 1 \,,
\end{equation}
 where $\Omega_{\varphi, d} = \bar\rho_{\varphi}/(\bar\rho_{r} + \bar\rho_{\varphi})$ is the dimensionless density parameter for the curvaton at the decay time $t_{d}$. Here, we take the sudden decay approximation for the curvaton, namely, the curvaton particles instantaneous decay into radiations.  And for simplicity, we will ignore the small curvature perturbation in the radiation fluid before the curvaton decays, i.e. $\zeta_{r}=0$. Then, order by order from Eq. (\ref{sudden rel}), we have 
 \begin{eqnarray}
	\zeta_{1} &=& f_{d} \zeta_{\varphi 1} \,, \label{transit} \\
	\zeta_{2} &=& \bigg[ \frac{3}{f_{d}} \left( \frac{2}{5}f_{NL}^{\varphi} +1\right) - 2 - f_{d} \bigg] \zeta_{1}^{2} \,, \\
	\zeta_{3} &=& \bigg[ \frac{54}{f_{d}^{2}} \left(  \frac{g_{NL}^{\varphi}}{25} +  \frac{f_{NL}^{\varphi}}{5} + \frac{1}{6} \right)
	-\frac{18}{f_{d}} \left( \frac{2}{5}f_{NL}^{\varphi} +1\right) -\frac{18}{5}f_{NL}^{\varphi}
	 -4+10f_{d} + 3f_{d}^{2}\bigg] \zeta_{1}^{3} \,,
\end{eqnarray}
where
\begin{equation}
	 f_{d} = \frac{3\Omega_{\varphi, d}}{4-\Omega_{\varphi, d}} \,, 
\end{equation}
is called the curvature perturbation transfer efficiency, see Ref.~\cite{curvaton}.  Therefore, the non-linearity parameters $f_{NL}$ and $g_{NL}$ are given by Eq.~(\ref{eqn:fnlgnl}) as
\begin{eqnarray}
	f_{NL} &=& \frac{1}{f_{d}} \left( f_{NL}^{\varphi} +\frac{5}{2}\right) - \frac{5(2 + f_{d})}{6}  \label{fdfgNL1}\,, \\
	g_{NL} &=&\frac{1}{f_{d}^{2}} \left(  g_{NL}^{\varphi} +  5f_{NL}^{\varphi} + \frac{25}{6} \right)
	-\frac{25}{3f_{d}} \left( \frac{2}{5}f_{NL}^{\varphi} +1\right) -\frac{5}{3}f_{NL}^{\varphi}
	 +\frac{25}{54}\bigg(10f_{d} + 3f_{d}^{2}-4\bigg) \,. \label{fdfgNL2}
\end{eqnarray}
In the limit of $f_{d}\rightarrow 1$, namely, the curvaton dominates the total energy density before it decays, we recover $f_{NL} \rightarrow f_{NL}^{\varphi}$ and $g_{NL} \rightarrow g_{NL}^{\varphi}$. While, in the limit of $f_{d}\rightarrow 0$, one can get large non-Gaussianity as ordinary curvaton models, but this is not the only way to get large non-Gassianity in the  $D-\lambda\varphi$  model, because one can get large non-linearity parameters of the curvaton itself $f_{NL}^{\varphi}$ as we mentioned before.

\subsection{The $L-\lambda\varphi$ model}
By using Eqs.~(\ref{fgNL1}), (\ref{fdfgNL1}) and (\ref{fdfgNL2}), we get
\begin{eqnarray}
	f_{NL} &=& \frac{5a_{1}}{2f_{d}} - \frac{5(2 + f_{d})}{6} \,, \\
	g_{NL} &=&\frac{25b_{1}}{6f_{d}^{2}}
	-\frac{25a_{1}}{3f_{d}}  -\frac{25a_{1}}{6}
	 +\frac{25(10f_{d} + 3f_{d}^{2})}{54}+ \frac{125}{54}  \label{L gnl}\,,
\end{eqnarray}
for the $L-\lambda\varphi$ model. Then, if $f_{d}\ll a_{1}$, $f_{NL}$ could be large, but $g_{NL}$ could be large or small depending on the exactly values of $a_{1}$, $b_{1}$ and $f_{d}$. Otherwise, if $a_{1}$, $b_{1}$ are much smaller than $f_{d}$, the terms with the factor $a_{1}$ or $b_{1}$ could be neglected,  then 
\begin{equation}\label{fgnl mod1}
	f_{NL} =  - \frac{5}{3}  - \frac{5 }{6}f_{d} \,, \quad g_{NL} = 25 \left(\frac{5f_{d}}{27}+ \frac{f_{d}^{2}}{18}+ \frac{5}{54} \right)\,.
\end{equation}
Thus, we have $ -5/2\leq f_{NL} <-5/3$ and $ 125/54< g_{NL}\leq25/3 $ for a Gaussian $\lambda$ field. And also we  have the relation $25 f_{NL} + 18g_{NL} = 0$ in the limit of $f_{d}\rightarrow 0$ in this situation.

\begin{figure}[h]
\begin{center}
\includegraphics[width=0.4\textwidth]{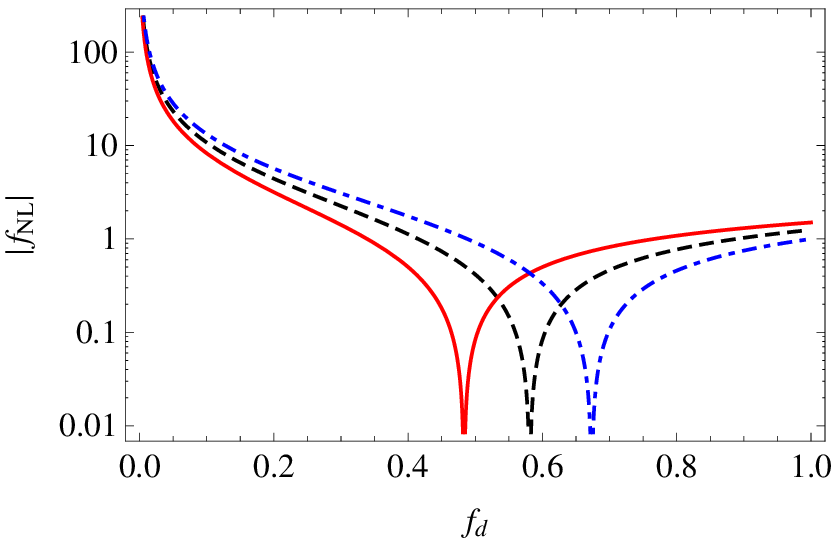}
\qquad
\includegraphics[width=0.4\textwidth]{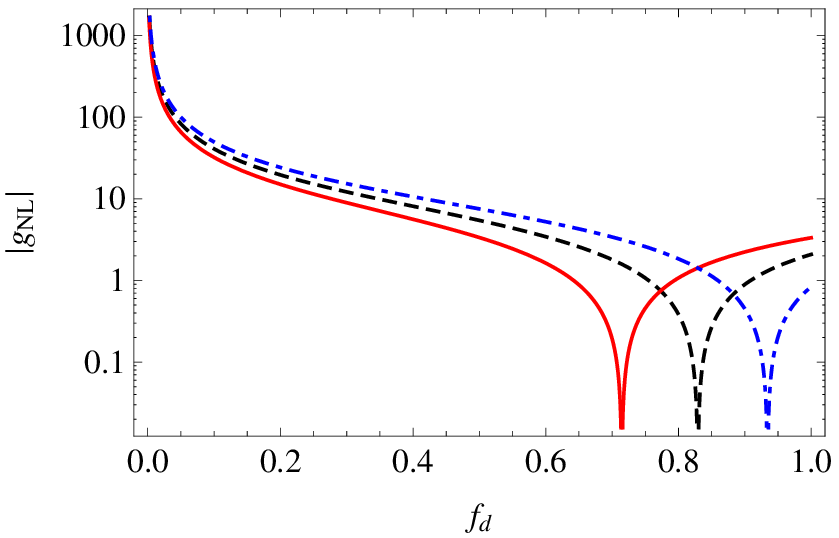}  
\caption{\label{fig::fnl1} The non-linearity parameters $f_{NL}$ (left) and $g_{NL}$ (right) as a function of the curvature perturbation transfer efficiency $f_{d}= \zeta_{1}/\zeta_{\varphi1}$ in the $L-\lambda\varphi$ model and the usual curvaton model (black dashed curve). The red solid curve corresponds to $a_{1}=0.4$, while the blue dot-dashed curve corresponds to $a_{1}=0.6$ and here we have set $b_{1}=0$. If $a_{1}=0.5, b_{1}=0 $, then the curves predicted in the $L-\lambda\varphi$ model and the usual curvaton model  are the same.}
\end{center}
\end{figure}

\begin{figure}[h]
\begin{center}
\includegraphics[width=0.4\textwidth]{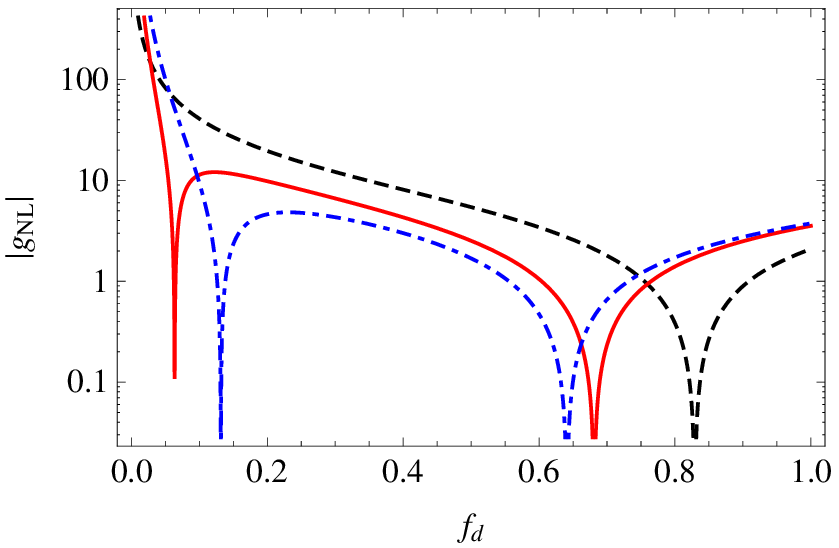}
\qquad
\includegraphics[width=0.4\textwidth]{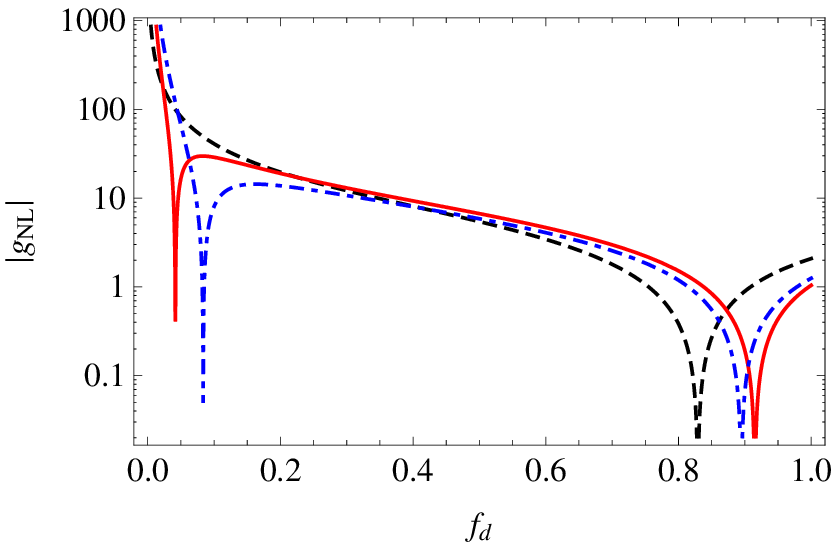}  
\caption{\label{fig::fnl12} The non-linearity parameters  $g_{NL}$ as a function of the curvature perturbation transfer efficiency $f_{d}= \zeta_{1}/\zeta_{\varphi1}$ with $a=0.4$(left) and $a=0.6$(right)  in the $L-\lambda\varphi$ model and the usual curvaton model (black dashed curve). The red solid curve corresponds to $b_{1}=0.05$, while the blue dot-dashed curve corresponds to $b_{1}=0.1$ in both figures.}
\end{center}
\end{figure}

To illustrate the above analytic study clearly, we plot the non-linearity parameters $f_{NL}$ and $g_{NL}$ as the function of $f_{d}$ (the transfer efficiency) in  Fig.~\ref{fig::fnl1} and Fig.~\ref{fig::fnl12}, and we also plot the results in a usual curvaton model \cite{curvaton} without nonlinear evolution of the curvaton field between the Hubble exit and the moment of its decay with a  black dashed curve in these figures.

From Fig.~\ref{fig::fnl1} , one can see that the shape of the curves in the $L-\lambda\varphi$ model are almost the same as that in the usual curvaton model. They would coincide together when  $a_{1}=0.5, b_{1}=0 $. If $b_{1}$ does not vanish, see Fig.~\ref{fig::fnl12}, the shape of the curves of $g_{NL}$  in the $L-\lambda\varphi$ model are much more different  with that in the usual curvaton model when $f_{d}\rightarrow 0$ due to the contribution of the first ($\sim f_{d}^{-2}$ ) terms in  Eq.~(\ref{L gnl}).

\subsection{The $D-\lambda\varphi$ model}

\begin{figure}[h]
\begin{center}
\includegraphics[width=0.4\textwidth]{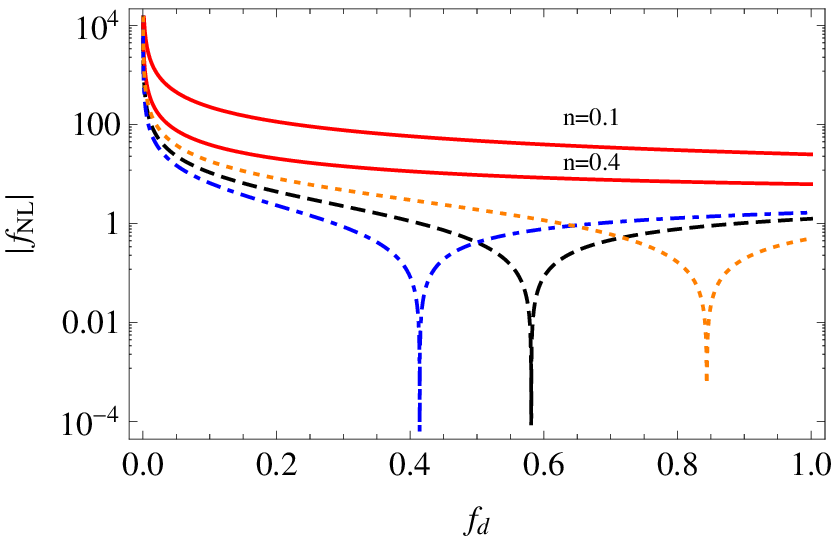}
\qquad
\includegraphics[width=0.4\textwidth]{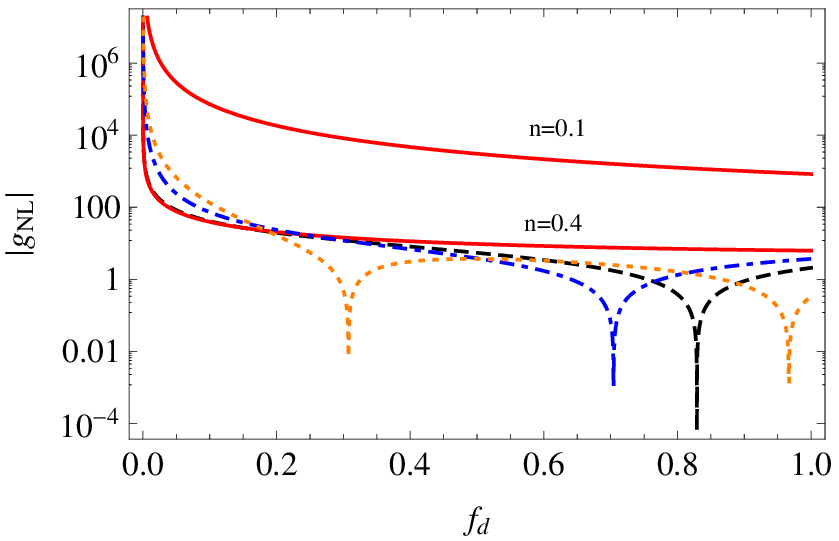}  
\caption{\label{fig::fnl2} The non-linearity parameters $f_{NL}$ (left) and $g_{NL}$ (right) as a function of the curvature perturbation transfer efficiency $f_{d}= \zeta_{1}/\zeta_{\varphi1}$  in the case of $\delta\lambda_{*} =0$ in the $D-\lambda\varphi$ model with $V\sim \varphi^{n}$ and the usual curvaton model (black dashed curve). The red solid curve corresponds to $n=0.1, 0.4$ from top to bottom, while the blue dot-dashed curve corresponds to $n=1.5$ and the orange dotted curve  corresponds to $n=5$.  If $n=2 $, then the curves in the $D-\lambda\varphi$ model will coincide  with that in the usual curvaton model. Here we have set $a_{3}=b_{3}=0$ for simplicity.}
\end{center}
\end{figure}

In the case of $\delta V_{*} =0$, the result is the same as that in the $L-\lambda\varphi$ model except for the definition of $a_{i}$ and $b_{i}$, so we will focus on the  case of   $\delta\lambda_{*} =0$ in this model. By using Eqs. (\ref{fgNl2}), (\ref{fdfgNL1}) and (\ref{fdfgNL2}), we get
\begin{eqnarray}
	f_{NL} &=& \frac{5}{2f_{d}} \left(a_{3} + \frac{ \bar V''_{*} \bar V_{*} } { \bar V_{*}^{'2} }\right) - \frac{5(2 + f_{d})}{6} \,, \\
	g_{NL} &=&\frac{25}{6f_{d}^{2}} \left(b_{3} + 3a_{3} \frac{ \bar V''_{*} \bar V_{*} } { \bar V_{*}^{'2} } + \frac{ \bar V'''_{*} \bar V_{*}^{2} } { \bar V_{*}^{'3} }  \right)
	-\left(\frac{25}{3f_{d}}+\frac{25}{6} \right) \left( a_{3} + \frac{ \bar V''_{*} \bar V_{*} } { \bar V_{*}^{'2} } \right)  +\frac{25(10f_{d} + 3f_{d}^{2})}{54}+ \frac{125}{54}
	  \,.
\end{eqnarray}
Now, letÕs forget about the $a_3$ and $b_3$ for a while, and take the function $V\sim\varphi^{n}$, then we have
\begin{eqnarray}
     f_{NL} &=&  \frac{5}{2f_{d}} \left(\frac{n-1}{n}\right)- \frac{5}{3}  - \frac{5 }{6}f_{d} \,, \\
     g_{NL} &=& 25 \left[\frac{(n-2)(n-1)}{6f_{d}^{2}n^{2}} - \left(\frac{1}{3f_{d}} +\frac{1}{6}\right)\left(\frac{n-1}{n}\right) + \frac{5f_{d}}{27}+ \frac{f_{d}^{2}}{18}+ \frac{5}{54} \right]\,,
\end{eqnarray}
which could be larger when $n$ is small. Of course, one can also get large non-Gaussianity  with $f_{d}\ll1$ when $n\neq1$. If $n=1$, then $f_{NL}$ and $g_{NL}$ have the same value as that in the $L-\lambda\varphi$ model, see Eq. (\ref{fgnl mod1}). If $n=2$ and in the limit of $f_{d}\rightarrow 0$,  we have $f_{NL} \rightarrow 5/(4f_{d})$, $g_{NL}\rightarrow -25/(6f_{d})$ and the relation $10f_{NL} + 3g_{NL} = 0$, which means $g_{NL}$ has the same order as $f_{NL}$. However, if $n\neq 1, 2$, we have
\begin{equation}
    f_{NL}\rightarrow \frac{5}{2f_{d}} \left(\frac{n-1}{n}\right) \,, \quad  g_{NL}\rightarrow \frac{25(n-2)(n-1)}{6f_{d}^{2}n^{2}} \,,
\end{equation}
in the limit of  $f_{d}\rightarrow 0$. And, this time 
\begin{equation}
   g_{NL} = \frac{2(n-2)}{3(n-1)} f_{NL}^{2} \,,
\end{equation}
which is much larger. The non-linearity parameters $f_{NL}$ and $g_{NL}$ as the function of $f_{d}$  are plotted in  Fig.~\ref{fig::fnl2} with $V\sim\varphi^{n}$,  and the results in a usual curvaton model  with linear evolution of the curvaton field between the Hubble exit and the moment of its decay are plotted with black dashed curves in these figures. From Fig.~\ref{fig::fnl2}, one can see that the shape of the curves are very similar except for a small value of $n$ with red solid curves, which means one can get large non-Gaussianity ($|f_{NL}|\gg1$) even if  the curvaton dominates the total energy density before it decays ($f_{d}\rightarrow1$) by setting a small $n$, e.g. $n=0.1$.

\begin{figure}[h]
\begin{center}
\includegraphics[width=0.4\textwidth]{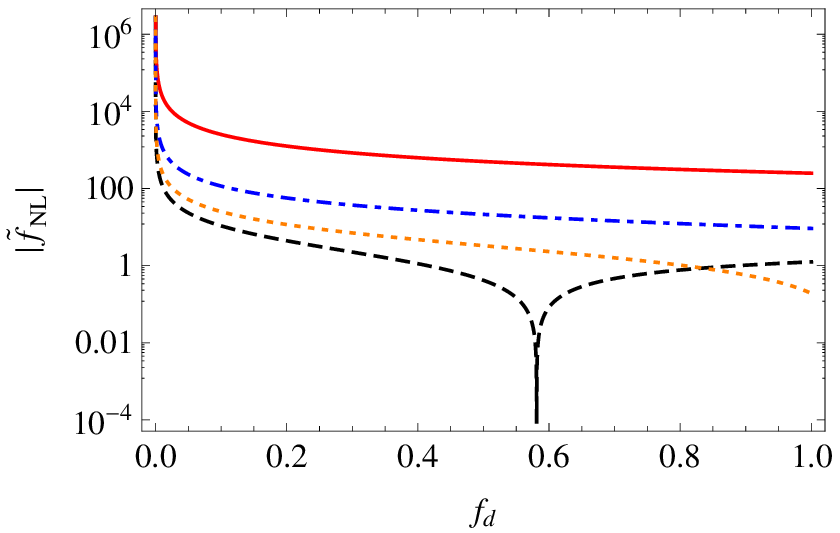}
\qquad
\includegraphics[width=0.4\textwidth]{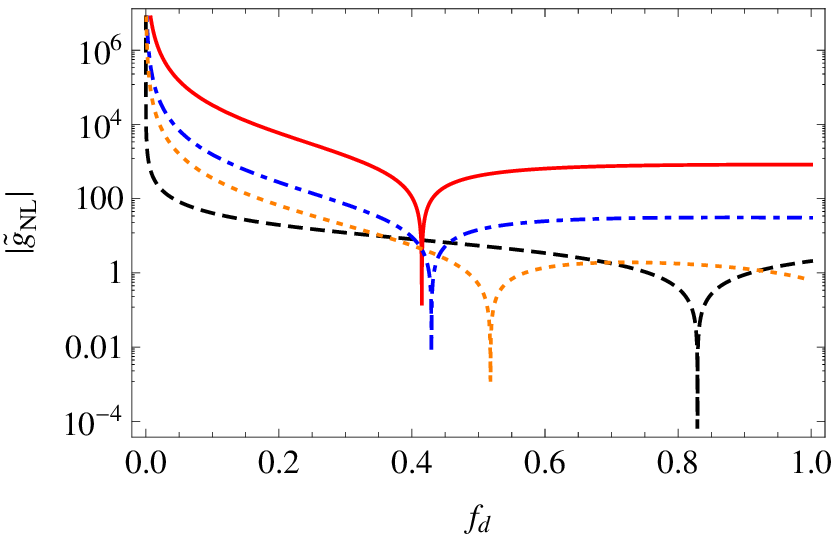}  
\caption{\label{fig::fnl3}The non-linearity parameters $f_{NL}$ (left) and $g_{NL}$ (right) as a function of the curvature perturbation transfer efficiency $f_{d}= \zeta_{1}/\zeta_{\varphi1}$  in the case of $\delta\lambda_{*} =0$ in the $D-\lambda\varphi$ model with $V\sim \cos(m\varphi)$ and the usual curvaton model (black dashed curve). The red solid curve corresponds to $m\varphi_{*}=0.1$, while the blue dot-dashed curve corresponds to $m\varphi_{*}=0.5$ and the orange dotted curve  corresponds to $m\varphi_{*}=2.0$.  Here we have set $a_{3}=b_{3}=0$ for simplicity.}
\end{center}
\end{figure}

If we take the function $V\sim\cosh(m\varphi)$, then we get
\begin{eqnarray}
	\tilde f_{NL} &=& \frac{5}{2f_{d} \tanh^{2}(m\varphi_{*})} - \frac{5(2 + f_{d})}{6}  \,, \\
	\tilde g_{NL} &=&25\bigg[\frac{1}{6f_{d}^{2} \tan^{2}(m\varphi_{*}) }
          - \left(\frac{1}{3f_{d}} +\frac{1}{6}\right) \frac{1}{ \tanh^{2}(m\varphi_{*})} 
	 +\frac{1}{54}\bigg(10f_{d} + 3f_{d}^{2}+5\bigg) \bigg] \,.
\end{eqnarray}
As we mentioned before, when $m\varphi_{*}\ll1$,  $\tilde f_{NL} $ and $\tilde g_{NL} $ could be larger even when $f_{d}=1$.  Taking the limit of $f_{d}\rightarrow0$, we get
\begin{equation}
       \tilde f_{NL} \rightarrow \frac{5}{2f_{d} \tanh^{2}(m\varphi_{*})}  \,, \quad \tilde g_{NL} \rightarrow \frac{25}{6f_{d}^{2} \tanh^{2}(m\varphi_{*}) } \,,
\end{equation}
 and the relation 
 \begin{equation}
      \tilde g_{NL}  = \frac{2}{3}\tanh^{2}(m\varphi_{*})\tilde f_{NL}^{2} \,.
\end{equation}
The non-linearity parameters $\tilde f_{NL}$ and $\tilde g_{NL}$ as the function of $f_{d}$  are plotted in  Fig.~\ref{fig::fnl3} with $V\sim\cosh(m\varphi)$. Again, the black dashed curves in these figures represent the results in a usual curvaton model  with linear evolution of the curvaton field between the Hubble exit and the moment of its decay, and the shape of the curves are not similar. For a small value of $m\varphi_{*}=0.1$ with red solid curves in the left figure,  one can get large non-Gaussianity ($|f_{NL}|\gg1$) even if  the curvaton dominates the total energy density before it decays ($f_{d}\rightarrow1$). 

\section{Probability Density Function}\label{sec:pdf}

In this section, we will  follow the method  in \cite{Sasaki:2006kq}  to calculate the probability density function (pdf) of curvature perturbation.  At first, we shall briefly review this method.  Let's assume there are two random variables $y$ and $z$, and the functional dependence of $z$ on $y$ is $z=z(y)$, which is  a bijection. If the pdf of $y$ is $\tilde f(y)$, then the probability of $z$ being in the interval $(z_{1},z_{2})$ is given by
\begin{equation}
	P(z_{1}<z<z_{2}) = \int_{z_{1}}^{z_{2}} \bigg|\frac{dy}{dz}\bigg|\tilde f(y) dz \,,
\end{equation}
 where the absolute value is need when $y(z)$ is a decreasing function. Hence, the pdf of $z$ is
 \begin{equation}
	f(z) = \bigg|\frac{dy(z)}{dz}\bigg|\tilde f[y(z)]  \,,
\end{equation}
where the derivative could be replaced by the Jacobian determinant in the multi-variable case.

Since the first order perturbation $\zeta_{1}$ only depends linearly on the initial Gaussian field perturbation,  one can take $\zeta_{1}$ as a Gaussian ``reference'' variable with mean $\mu_{\zeta_{1}} = 0$. In the sudden decay approximation, we have found an analytic functional dependence $\zeta=\zeta(\zeta_{1})$, but the mapping is not always  a bijection. Calling these values $\zeta_{1i}$, one can calculate the pdf of the non-linear primordial curvature perturbation   
\begin{equation}
	f(\zeta) = \sum_{i} \bigg|\frac{d\zeta_{1}}{d\zeta}\bigg|_{\zeta_{1}=\zeta_{1i}} f_{g}(\zeta_{1i}) \,,
\end{equation}
 where $f(\zeta_{1})$ is the Gaussian pdf with mean $\mu=0$ and variance $\sigma^{2} = \sigma^{2}_{\zeta_{1}}$:
 \begin{equation}
	f_{g}(\zeta_{1}) = \frac{1}{\sqrt{2\pi \sigma^{2}_{\zeta_{1}}}} e^{-\zeta_{1}^{2}/(2\sigma_{\zeta_{1}}^{2})}  \,.
\end{equation}
For simplicity, in the rest of this section, we will neglect all the non-linear fluctuation of the initial field, namely, we will set $a_{i}=b_{i}=0$ in all models. 

Actually, the non-Gaussianity could be described quantitatively by calculating the moments of the pdf and 
\begin{equation}
       m_{z}(i) = \int (z-\mu)^{i} f(z) dz \,,
\end{equation}
 is called the $i^{\text{th}}$ moment. Here, the mean $\mu$ can be calculated as 
 \begin{equation}
      \mu = \int z f(z) dz \,,
\end{equation}
where the pdf $f(z)$ satisfies $\int f(z)dz = 1$. Conventionally, the second moment is the variance ($\sigma^{2}_{z}$), the third moment is called skewness, and the fourth moment kurtosis.  For a Gaussian pdf, any odd moment (with $i\ge3$) is zero, since the probability density is symmetric around the mean, while the even moments could be easily calculated by partial integrating, e.g. $m(4)=3\sigma^{4}$, $m(6)=15\sigma^{6}$, $m(8)=105\sigma^{8}$, $m(10)=945\sigma^{10}$, $m(12)=10395\sigma^{12}$, etc.,  see Ref.~\cite{Sasaki:2006kq}.  Any departure from these values indicates the pdf is non-Gaussian, namely, there is an asymmetric deviation from Gaussianity, if odd moments are not zero, and if the pdf is more (or less) sharply  peaked than the Gaussian if  even moments are smaller (or larger) than that in the Gaussian case. Therefore, the set of moments encodes the same information  of non-Gaussianity as the fully non-linear $\zeta$ or its expansion. Next, we will calculate the pdf and illustrate some moments of the pdf in our models.

\subsection{The $L-\lambda\varphi$ model}
From Eqs.~(\ref{equ:flat fluc 1}), (\ref{equ:rho1}), (\ref{equ:zeta1}) and (\ref{transit}), we have
\begin{equation}\label{equ:e31}
	e^{3\zeta_{\varphi}} = 1+\frac{3\zeta_{1}}{f_{d}} \,.
\end{equation}
 While, from Eq.~(\ref{sudden rel}) with $\zeta_{r}=0$, we have
\begin{equation}\label{equ:e32}
        e^{3\zeta_{\varphi}} = \frac{3+f_{d}}{4f_{d}} e^{3\zeta} + \frac{3f_{d}-3}{4f_{d}} e^{-\zeta} \,.
\end{equation}
Thus, combing  Eqs.~(\ref{equ:e31}) and (\ref{equ:e32}), we get
\begin{equation}
        \zeta_{1} = \frac{f_{d}}{3} \left(-1+\frac{3+f_{d}}{4f_{d}} e^{3\zeta} + \frac{3f_{d}-3}{4f_{d}} e^{-\zeta} \right) \,,
\end{equation}
and 
\begin{equation}
	\bigg| \frac{d\zeta_{1}}{d\zeta} \bigg| = \frac{3+f_{d}}{4} e^{3\zeta} + \frac{1-f_{d}}{4} e^{-\zeta}  \,.
\end{equation}
 Hence, the non-Gaussian probability density function for $\zeta$ is 
 \begin{equation}
	f(\zeta) = \frac{1}{\sqrt{2\pi \sigma^{2}_{\zeta_{1}}}} \left(\frac{3+f_{d}}{4} e^{3\zeta} + \frac{1-f_{d}}{4} e^{-\zeta}\right)
	 \exp\bigg[{-\frac{f^{2}_{d}}{9} \left(-1+\frac{3+f_{d}}{4f_{d}} e^{3\zeta} + \frac{3f_{d}-3}{4f_{d}} e^{-\zeta} \right) ^{2}/(2\sigma_{\zeta_{1}}^{2})} \bigg] \,.
 \end{equation}

\begin{figure}[h]
\begin{center}
\includegraphics[width=0.4\textwidth]{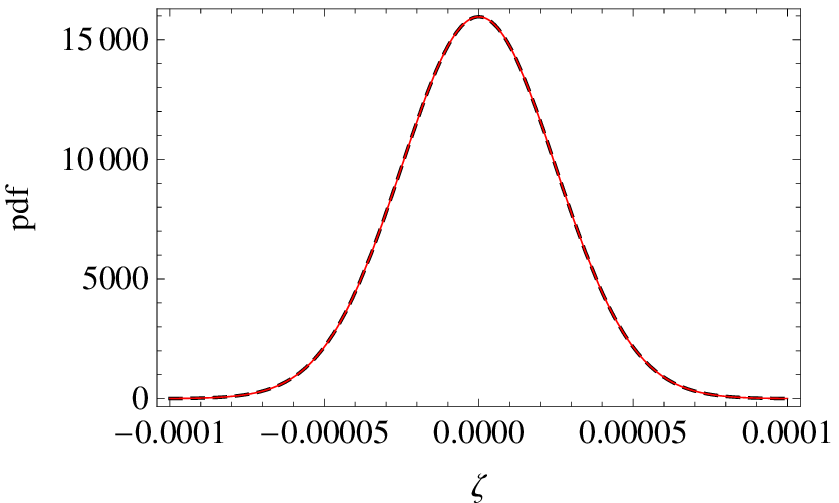}
\qquad
\includegraphics[width=0.4\textwidth]{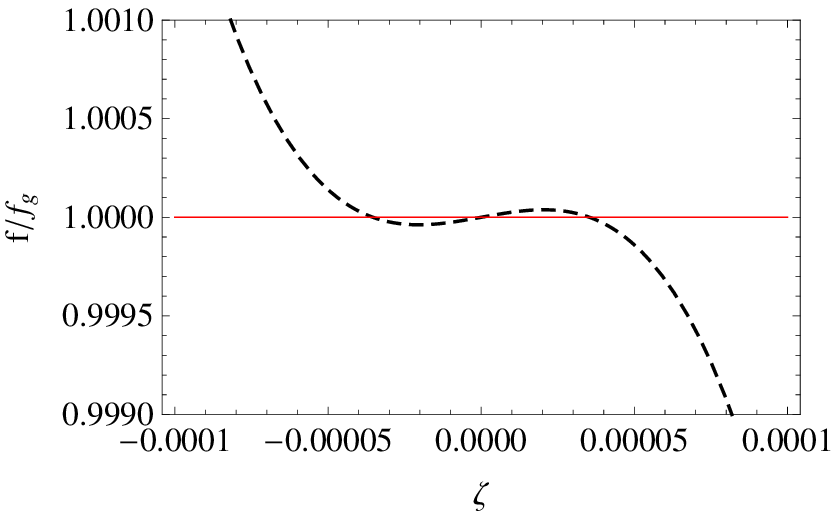}  
\caption{\label{fig::pdf1} \textit{The $L-\lambda\varphi$ model.} Left: Pdfs at $f_{d}=0.8$, ($f_{NL}=-7/3$). The black dashed curve is the pdf $f$, while the red solid curve is the Gaussian reference, $f_{g}$. Right: The ratio of non-Gaussian pdf to the Gaussian one.  }
\end{center}
\end{figure} 

\begin{figure}[h]
\begin{center}
\includegraphics[width=0.4\textwidth]{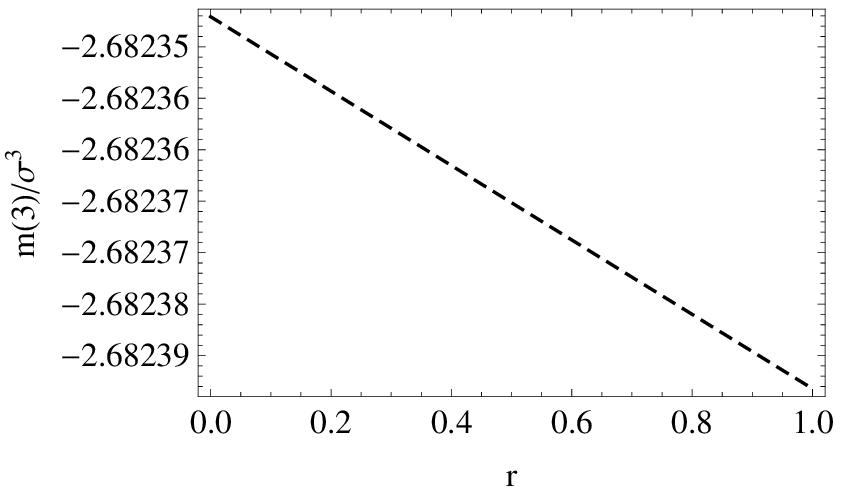}
\qquad
\includegraphics[width=0.4\textwidth]{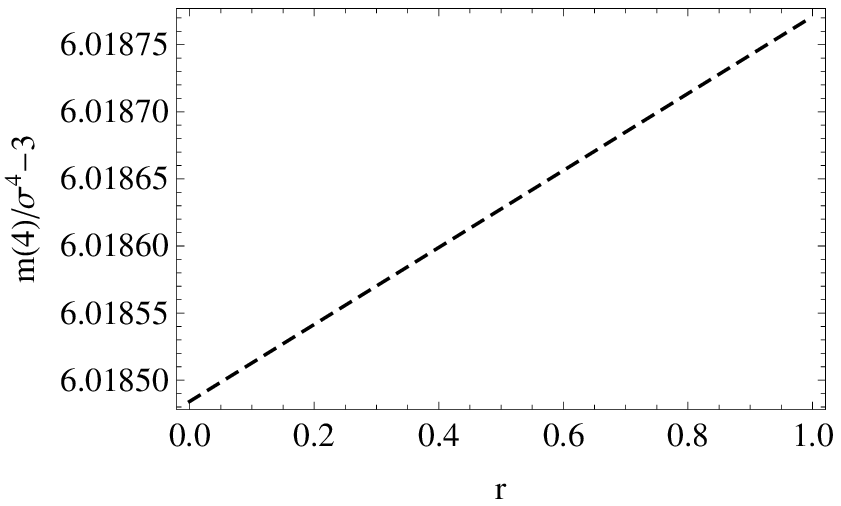}  
\qquad
\includegraphics[width=0.4\textwidth]{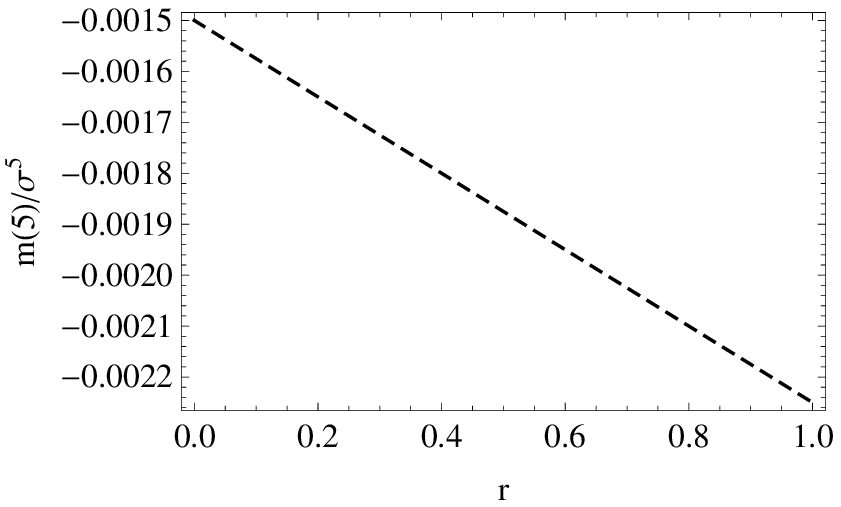}  
\qquad
\includegraphics[width=0.4\textwidth]{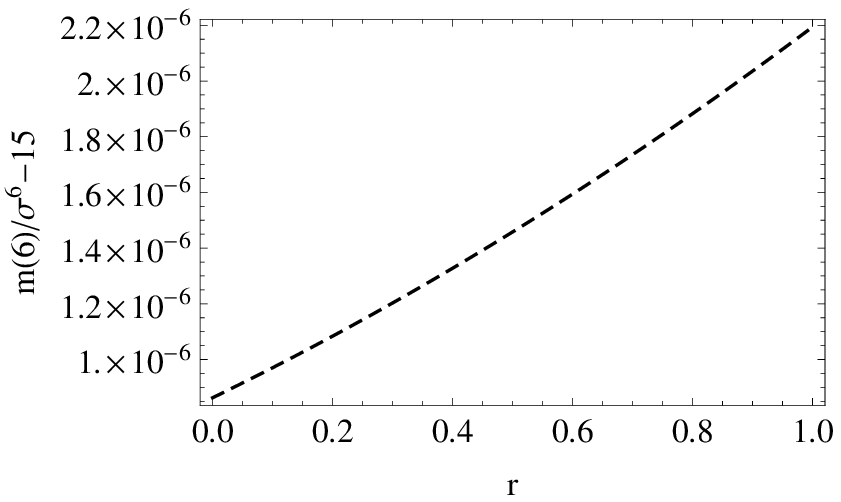}  
\caption{\label{fig::pdfm1} Third, fourth, fifth and sixth (from left top to right bottom) moments of the pdf of the primordial curvature perturbation $\zeta$ as a function of the linear transfer parameter, $f_{d}$ in the $L-\lambda\varphi$ model.}
\end{center}
\end{figure} 

In Fig.~\ref{fig::pdf1}, we compare the fully non-linear pdf $f(\zeta)$ to the Gaussian $f_{g}(\zeta_{1})$, and it shows that $f$ is virtually indistinguishable from the Gaussian $f_{g}$. But, we also plot $f/f_{g}$ which reveals the non-Gaussianity in this model. In Fig.~\ref{fig::pdfm1}, we plot the moments from the third up to the sixth one as a function of $f_{d}$.

\subsection{The $D-\lambda\varphi$ model}

In this model, we will focus on the case of   $\delta\lambda_{*} =0$.  From Eq.~(\ref{equ:flat fluc 1}), (\ref{equ:rho2}) and (\ref{equ:zeta13}), we have
\begin{equation}\label{equ:e33}
	e^{3\zeta_{\varphi}} = 1+3\zeta_{\varphi1}  + \frac{9}{2} \frac{\bar V''_{*} V_{*}}{ V^{'2}_{*}}(\zeta_{\varphi1})^{2} 
	+ \frac{9}{2} \frac{\bar V'''_{*} V^{2}_{*}}{ V^{'3}_{*}}(\zeta_{\varphi1})^{3} \,.
\end{equation}
and by combing Eq.~(\ref{equ:e32}), we can get the equation for $\zeta_{1}$ for a given function $V$. Next, we will consider  two types of $V$ and get the  non-Gaussian probability density function for $\zeta$  by solving the equation as in the $L-\lambda\varphi$ model.

\subsubsection{$V\sim\varphi^{n}$}

In this case  we get the following equation for $\zeta_{\varphi1}$:
\begin{equation}
	\frac{9(n-1)(n-2)}{2n^{2}} \left(\frac{\zeta_{1}}{f_{d}} \right)^{3} + \frac{9(n-1)}{2n}\left (\frac{\zeta_{1}}{f_{d}}\right)^{2} +3\left(\frac{\zeta_{1}}{f_{d}} \right)+ d= 0 \,,
\end{equation}
where 
\begin{equation}
      d = 1 - \frac{3+f_{d}}{4f_{d}} e^{3\zeta} -\frac{3f_{d}-3}{4f_{d}} e^{-\zeta} \,.
\end{equation}
Here we have used Eqs.~(\ref{equ:e32}), (\ref{equ:e33}) and (\ref{transit}). If $n=1$, the results would be the same as that in the $L-\lambda\varphi$ model, while if $n=2$, we recover the results of the usual curvaton model. For a generic $n$ ($n\neq 1,2$), 
the above equation could be rewritten as
\begin{equation}\label{equ:Y}
       Y^{3} + p Y + q = 0 \,,
\end{equation}
where we have defined  
\begin{eqnarray}
      Y &=& \frac{\zeta_{1}}{f_{d}} + \frac{n}{3(n-2)} \,,  \\
      p &=&  \frac{n^{2}(n-3)}{3(n-2)^{2}(n-1)} \,,\\
      q &=& \frac{2n^{2}}{9(n-2)} \left( \frac{d }{n-1} - \frac{n}{(n-1)(n-2)} +\frac{n}{3(n-2)^{2}} \right) \,.
\end{eqnarray}
Define 
\begin{equation}
     D= \left( \frac{p}{3}\right)^{3} + \left(\frac{q}{2} \right)^{2} \,,
\end{equation}
then, the number of real roots of the Eq. (\ref{equ:Y}) depends on the sign of $D$. It should be noticed that if $n>3$ or $n<1$, $p$ is positive and then $D$ is also positive. So, in this case, we have only one real root of the Eq. (\ref{equ:Y}):
\begin{equation}
    Y_{1} = u_{+}^{1/3} + u_{-}^{1/3}  \,, \quad u_{\pm} = \frac{-q}{2} \pm D^{1/2} \,,
\end{equation}
and 
\begin{equation}
	\bigg| \frac{d\zeta_{1}}{d\zeta} \bigg| =  \frac{Q}{2\sqrt{D}} 
	\bigg|u_{-}^{1/3}-u_{+}^{1/3}\bigg|\,.
\end{equation}
where 
\begin{equation}
      Q =  \frac{n^{2}\big[(3+f_{d})e^{3\zeta} + (1-f)e^{-\zeta}\big]}{18 |n-1||n-2|}\,.
\end{equation}
 Hence, the non-Gaussian probability density function for $\zeta$ is 
 \begin{equation}
	f(\zeta)\Big|_{D>0} = \frac{1}{\sqrt{2\pi \sigma^{2}_{\zeta_{1}}}} \left(  \frac{Q\big |u_{-}^{1/3}-u_{+}^{1/3}\big| }{2\sqrt{D}} 
	\right)
	 \exp\bigg[{-\frac{f^{2}_{d}}{2\sigma_{\zeta_{1}}^{2}} \left(u_{+}^{1/3} + u_{-}^{1/3} -  \frac{n}{3(n-2)}  \right) ^{2}} \bigg]  \,.
\end{equation}

If  $ 1<n<3$ and $n\neq2$, there could be some regions $\zeta \in(\zeta_{a}, \zeta_{b})$, in which $D< 0$ and $D|_{\zeta=\zeta_{a} } =D|_{\zeta=\zeta_{b} } =0 $. So, for completeness, we will give the pdf for $\zeta$ in these regions in the following. If $D=0$, we have three  real roots of the Eq. (\ref{equ:Y}) and two of them are equal:
\begin{equation}
    Y_{1} = 2 \left(\frac{-q}{2}\right)^{1/3}  \,, \quad
    Y_{2} = Y_{3} = \left(\frac{q}{2} \right)^{1/3}  \,,
\end{equation}
and then, we have
\begin{equation}
    \bigg| \frac{d\zeta^{(1)}_{1}}{d\zeta} \bigg| = Q\bigg|\left(\frac{-q}{2}\right)^{-2/3}\bigg|\,,\quad
    \bigg| \frac{d\zeta^{(2)}_{1}}{d\zeta} \bigg| =  \frac{Q}{2} \bigg|\left(\frac{q}{2}\right)^{-2/3}\bigg|\,.
\end{equation}
 Hence, the non-Gaussian probability density function for $\zeta$ is 
 \begin{eqnarray}
 \nonumber
f(\zeta)\Big|_{D=0} &=& \frac{Q}{\sqrt{2\pi \sigma^{2}_{\zeta_{1}}}} \Bigg\{
	  \bigg|\left(\frac{-q}{2}\right)^{-2/3}\bigg|
	 \exp\bigg[{-\frac{f^{2}_{d}}{2\sigma_{\zeta_{1}}^{2}} \left(2 \left(\frac{-q}{2}\right)^{1/3}  -  \frac{n}{3(n-2)}  \right) ^{2}} \bigg] \\
	 &+&\frac{1}{2} \bigg|\left(\frac{q}{2}\right)^{-2/3}\bigg|
	 \exp\bigg[{-\frac{f^{2}_{d}}{2\sigma_{\zeta_{1}}^{2}} \left(\left(\frac{q}{2}\right)^{1/3}  -  \frac{n}{3(n-2)}  \right) ^{2}} \bigg]
	 \Bigg\} \,.
\end{eqnarray}
And, if $D<0$, we have three different real roots of the Eq. (\ref{equ:Y}):
\begin{equation}
    Y_{1} = 2\sqrt{\frac{|p|}{3}} \cos\frac{\theta}{3} \,, \quad Y_{2,3} = - 2\sqrt{\frac{|p|}{3}} \cos \frac{\theta\pm\pi}{3} \,,
\end{equation}
where 
\begin{equation}
    \theta = \arccos\left[ \frac{-q}{2} \left(\frac{|p|}{3} \right)^{-3/2}\right] \,.
\end{equation}
Thus
\begin{eqnarray}
    \bigg| \frac{d\zeta^{(1)}_{1}}{d\zeta} \bigg| &=& Q \sin \frac{\theta}{3} \sin^{-1}\theta \left(\frac{|p|}{3}\right)^{-1}\,, \\
    \bigg| \frac{d\zeta^{(2,3)}_{1}}{d\zeta} \bigg| &=&  Q \sin\frac{\theta\pm\pi}{3}\sin^{-1}\theta \left(\frac{|p|}{3}\right)^{-1}\,. 
\end{eqnarray}
 Hence, the non-Gaussian probability density function for $\zeta$ is 
 \begin{eqnarray}
 \nonumber
f(\zeta)\Big|_{D<0} &=& \frac{3Q}{\sqrt{2\pi \sigma^{2}_{\zeta_{1}}}|p\sin\theta| }\Bigg\{
	\left|\sin\frac{\theta}{3}\right|
	 \exp\bigg[{-\frac{f^{2}_{d}}{2\sigma_{\zeta_{1}}^{2}} \left(2\sqrt{\frac{|p|}{3}} \cos\frac{\theta}{3}-  \frac{n}{3(n-2)}  \right) ^{2}} \bigg] \\
	 &+& \sum_{\pm}\left|\sin\frac{\theta\pm\pi}{3}\right|
	 \exp\bigg[{-\frac{f^{2}_{d}}{2\sigma_{\zeta_{1}}^{2}} \left(2\sqrt{\frac{|p|}{3}} \cos\frac{\theta\pm\pi}{3} +  \frac{n}{3(n-2)}  \right) ^{2}} \bigg]
	 \Bigg\} \,.
\end{eqnarray}
 \begin{figure}[h]
\begin{center}
\includegraphics[width=0.4\textwidth]{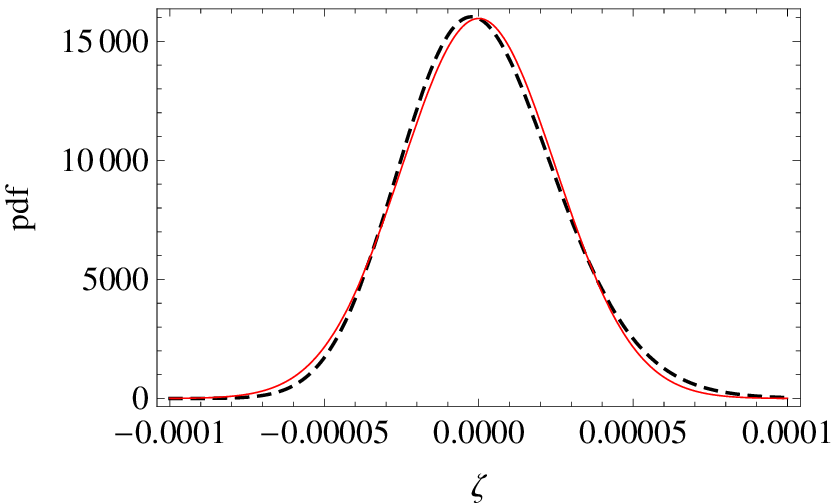}
\qquad
\includegraphics[width=0.4\textwidth]{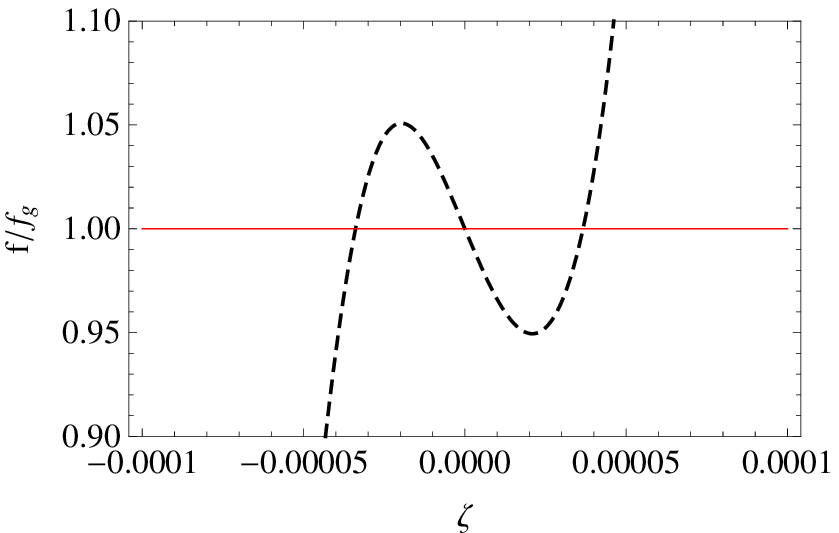}  
\qquad
\includegraphics[width=0.4\textwidth]{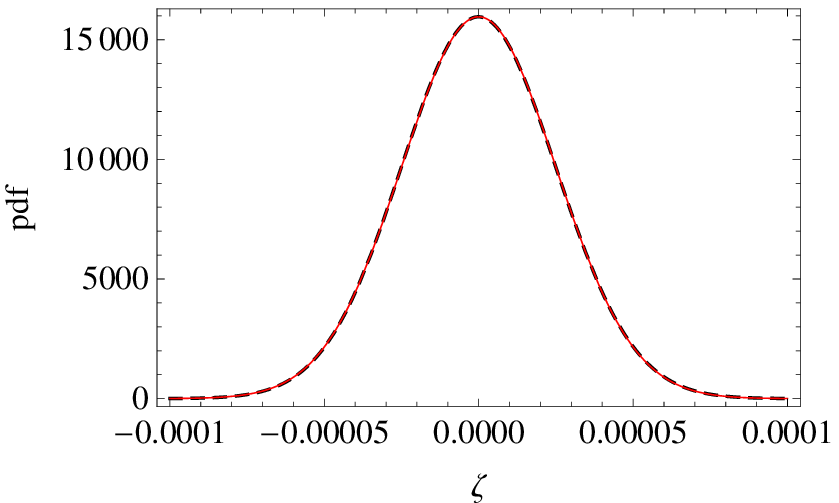}  
\qquad
\includegraphics[width=0.4\textwidth]{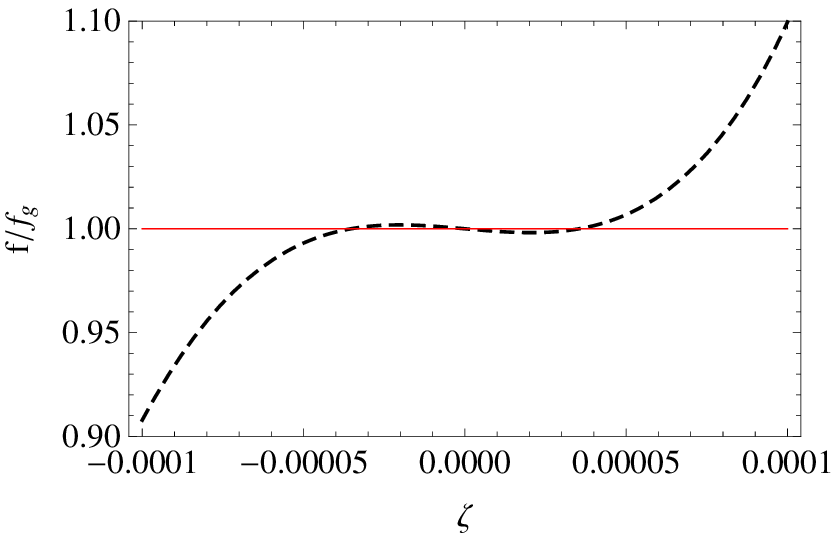}  
\caption{\label{fig::pdf2}\textit{The $D-\lambda\varphi$ model with $V\sim\varphi^{n}$.} Left:  Pdfs at $f_{d}=0.8$ and $n=-0.001$ (top, $f_{NL}=3126$), $-0.0276$ (bottom, $f_{NL}=114$). The black dashed curve ares the pdfs $f$, while the red solid curves are the Gaussian reference, $f_{g}$. Right: The ratio of non-Gaussian pdf to the Gaussian one with $n=-0.001$ (top), $-0.0276$ (bottom). }
\end{center}
\end{figure} 

\begin{table}[h]
\centering
  \begin{tabular}{l|l|l|l|l|l|l}
  \hline
  \hline
  \multirow{2}{*}{Moment} &  \multicolumn{3}{c|}{$n=-0.0010$} & \multicolumn{3}{c}{$n=-0.0276$}                   	\\
  \cline{2-7}
  & \multicolumn{1}{c|}{$f_{d} = 0.2$} & \multicolumn{1}{c|}{$f_{d} = 0.5$}  & \multicolumn{1}{c|}{$f_{d} = 0.8$} 
  & \multicolumn{1}{c|}{$f_{d} = 0.2$} & \multicolumn{1}{c|}{$f_{d} = 0.5$}  & \multicolumn{1}{c}{$f_{d} = 0.8$}  	\\
  \hline
  \hline
  $m(3)/\sigma^{3}$      & $-6.9688 \times 10^{-3} $ & $-3.5883 \times 10^{-3} $ & $-2.3228 \times 10^{-3} $
  				     & $-3.5297 \times 10^{-4} $ & $-1.4023 \times 10^{-4} $ & $-8.6818 \times 10^{-5} $   	\\
  \hline
  $m(4)/\sigma^{4}-3$   & $-2.9970$                         &$-2.9980$                          & $-2.9982$            
  				     & $-2.9985$                         &$-2.9985$                          & $-2.9985$                               	\\
  \hline
  $m(5)/\sigma^{5}$      & $-1.2190 \times 10^{-3}$  & $-5.8153 \times 10^{-4} $ & $-3.6840 \times 10^{-4}$              
  				     & $-5.5229 \times 10^{-5}$  & $-2.1891 \times 10^{-5} $ & $-1.3559 \times 10^{-5} $	   	\\
  \hline
  $m(6)/\sigma^{6}-15$ & $-14.999$                         & $-14.999$                         & $-14.999$                            
  				     & $-14.999$                         & $-14.999$                         & $-14.999$                                 	\\
  \hline
  \hline
  \end{tabular}
  \caption{\label{table:moment2}  The value of  the third, fourth, fifth and sixth  moments of the pdf of the primordial curvature perturbation $\zeta$ with $n=-0.001, -0.0276$ and $f_{d}=0.2, 0.5, 0.8$ in the $D-\lambda\varphi$ model with  $\delta\lambda_{*} =0$ and $V\sim\varphi^{n}$. }
\end{table}

In Fig.~\ref{fig::pdf2}, we compare the fully non-linear pdf $f(\zeta)$ to the Gaussian $f_{g}(\zeta_{1})$ in the case when there is  large non-Gaussianity. In one of them, the non-linearity parameter is very large ($f_{d}=0.8$, $n=-0.001$, $f_{NL}=3126$), and this kind of visual comparison reveals the non-Gaussianity, but in another one with  ($f_{d}=0.8$, $n=-0.0276$, $f_{NL}=114$),  it shows that $f$ is virtually indistinguishable from the Gaussian $f_{g}$. But, we also plot $f/f_{g}$ which reveals the non-Gaussianity in this model. In Tab.~\ref{table:moment2}, we give some values of the moments from the third up to the sixth one at $f_{d} = 0.2, 0.5, 0.8$ for $n=-0.001$ and $n=-0.0267$.

\subsubsection{$V\sim\cosh(m\varphi)$}

In this case,  we get the following equation for $\zeta_{\varphi1}$ from Eq.~(\ref{equ:e32}) and (\ref{equ:e33}):
 \begin{equation}
	\left(\frac{\zeta_{1}}{f_{d}} \right)^{3} +\left (\frac{\zeta_{1}}{f_{d}}\right)^{2} +\frac{2 \tanh^{2}(m\varphi_{*})}{3}\left(\frac{\zeta_{1}}{f_{d}} \right)+ \frac{2d\tanh^{2}(m\varphi_{*})} {9}= 0 \,,
\end{equation}
which could be rewritten as
\begin{equation}\label{equ:Z}
        Z^{3} + \tilde p Z + \tilde q = 0 \,,
\end{equation}
where we have defined  
\begin{equation}
	      Z = \frac{\zeta_{1}}{f_{d}} + \frac{1}{3} \,,  \quad
      \tilde p =  \frac{2\tanh^{2}(m\varphi_{*})-1}{3} \,,\quad
      \tilde q = \frac{2}{27} + \frac{2(d-1)}{9}\tanh^{2}(m\varphi_{*})\,.
\end{equation}
Thus, the solutions to Eq.~(\ref{equ:Z}) are the same as that to Eq.~(\ref{equ:Y}) except for the definitions of the coefficients $p$ and $q$. Define 
\begin{equation}
     \tilde D= \left( \frac{\tilde p}{3}\right)^{3} + \left(\frac{\tilde q}{2} \right)^{2} \,,
\end{equation}
whose sign determines the number of roots of the Eq.~(\ref{equ:Z}).  It should be noticed that if $\tan^{2}(m\varphi_{*})>1/2$, $\tilde p$ is positive and then $\tilde D$ is also positive. So, in this case, we have only one real root of the Eq. (\ref{equ:Z}):
\begin{equation}
    Z_{1} = \tilde u_{+}^{1/3} +\tilde u_{-}^{1/3}  \,, \quad \tilde u_{\pm} = \frac{-\tilde q}{2} \pm\tilde D^{1/2} \,,
\end{equation}
and 
\begin{equation}
	\bigg| \frac{d\zeta_{1}}{d\zeta} \bigg| =  \frac{\tilde Q}{2\sqrt{\tilde D}} 
	\bigg|\tilde u_{-}^{1/3}-\tilde u_{+}^{1/3}\bigg|\,.
\end{equation}
where 
\begin{equation}
      \tilde Q =  \frac{\tanh^{2}(m\varphi_{*})}{18}\bigg[(3+f_{d})e^{3\zeta} + (1-f)e^{-\zeta}\bigg]\,.
\end{equation}
 Hence, the non-Gaussian probability density function for $\zeta$ is 
 \begin{equation}
	f(\zeta)\Big|_{\tilde D>0} = \frac{1}{\sqrt{2\pi \sigma^{2}_{\zeta_{1}}}} \left(  \frac{\tilde Q\big |\tilde u_{-}^{1/3}-\tilde u_{+}^{1/3}\big| }{2\sqrt{\tilde D}} 
	\right)
	 \exp\bigg[{-\frac{f^{2}_{d}}{2\sigma_{\zeta_{1}}^{2}} \left(\tilde u_{+}^{1/3} + \tilde u_{-}^{1/3} -  \frac{1}{3}  \right) ^{2}} \bigg]  \,.
\end{equation}

If  $ \tan^{2}(m\varphi_{*})<1/2$, there could be some regions  of $\zeta \in(\zeta_{a}, \zeta_{b})$, in which $\tilde D< 0$ and $\tilde D|_{\zeta=\zeta_{a} } =\tilde D|_{\zeta=\zeta_{b} } = 0 $. So, for completeness, we will give the pdf for $\zeta$ in these regions in the following. If $\tilde D=0$, we have three  real roots of the Eq. (\ref{equ:Z}) and two of them are equal:
\begin{equation}
    Z_{1} = 2 \left(\frac{-\tilde q}{2}\right)^{1/3}  \,, \quad
    Z_{2} = Y_{3} = \left(\frac{\tilde q}{2} \right)^{1/3}  \,,
\end{equation}
and then, we have
\begin{equation}
    \bigg| \frac{d\zeta^{(1)}_{1}}{d\zeta} \bigg| = \tilde Q\bigg|\left(\frac{-\tilde q}{2}\right)^{-2/3}\bigg|\,,\quad
    \bigg| \frac{d\zeta^{(2)}_{1}}{d\zeta} \bigg| =  \frac{\tilde Q}{2} \bigg|\left(\frac{\tilde q}{2}\right)^{-2/3}\bigg|\,.
\end{equation}
 Hence, the non-Gaussian probability density function for $\zeta$ is 
 \begin{eqnarray}
 \nonumber
f(\zeta)\Big|_{\tilde D=0} &=& \frac{\tilde Q}{\sqrt{2\pi \sigma^{2}_{\zeta_{1}}}} \Bigg\{
	  \bigg|\left(\frac{-\tilde q}{2}\right)^{-2/3}\bigg|
	 \exp\bigg[{-\frac{f^{2}_{d}}{2\sigma_{\zeta_{1}}^{2}} \left(2 \left(\frac{-\tilde q}{2}\right)^{1/3}  - \frac{1}{3}  \right) ^{2}} \bigg]  \\
	 &+&\frac{1}{2} \bigg|\left(\frac{\tilde q}{2}\right)^{-2/3}\bigg|
	 \exp\bigg[{-\frac{f^{2}_{d}}{2\sigma_{\zeta_{1}}^{2}} \left(\left(\frac{\tilde q}{2}\right)^{1/3}  -  \frac{1}{3}  \right) ^{2}} \bigg]
	 \Bigg\} \,.
\end{eqnarray}
And, for $\tilde D<0$, we have three different real roots of the Eq. (\ref{equ:Z}):
\begin{equation}
    Z_{1} = 2\sqrt{\frac{|\tilde p|}{3}} \cos\frac{\tilde \theta}{3} \,, \quad Z_{2,3} = - 2\sqrt{\frac{|\tilde p|}{3}} \cos \frac{\tilde \theta\pm\pi}{3} \,,
\end{equation}
where 
\begin{equation}
    \tilde \theta = \arccos\left[ \frac{-\tilde q}{2} \left(\frac{|\tilde p|}{3} \right)^{-3/2}\right] \,.
\end{equation}
Thus
\begin{eqnarray}
    \bigg| \frac{d\zeta^{(1)}_{1}}{d\zeta} \bigg| &=& \tilde Q \sin \frac{\tilde \theta}{3} \sin^{-1}\tilde \theta \left(\frac{|\tilde p|}{3}\right)^{-1}\,, \\
    \bigg| \frac{d\zeta^{(2,3)}_{1}}{d\zeta} \bigg| &=&  \tilde Q \sin\frac{\tilde \theta\pm\pi}{3}\sin^{-1}\tilde \theta \left(\frac{|\tilde p|}{3}\right)^{-1}\,. 
\end{eqnarray}
 Hence, the non-Gaussian probability density function for $\zeta$ is 
 \begin{eqnarray}
 \nonumber
f(\zeta)\Big|_{\tilde D<0} &=& \frac{3\tilde Q}{\sqrt{2\pi \sigma^{2}_{\zeta_{1}}}|\tilde p\sin\tilde \theta| }\Bigg\{
	\left|\sin\frac{\tilde \theta}{3}\right|
	 \exp\bigg[{-\frac{f^{2}_{d}}{2\sigma_{\zeta_{1}}^{2}} \left(2\sqrt{\frac{|\tilde p|}{3}} \cos\frac{\tilde \theta}{3}- \frac{1}{3}  \right) ^{2}} \bigg] \\
	 &+& \sum_{\pm}\left|\sin\frac{\tilde \theta\pm\pi}{3}\right|
	 \exp\bigg[{-\frac{f^{2}_{d}}{2\sigma_{\zeta_{1}}^{2}} \left(2\sqrt{\frac{|\tilde p|}{3}} \cos\frac{\tilde \theta\pm\pi}{3} +  \frac{1}{3}  \right) ^{2}} \bigg]
	 \Bigg\} \,.
\end{eqnarray}

 \begin{figure}[h]
\begin{center}
\includegraphics[width=0.4\textwidth]{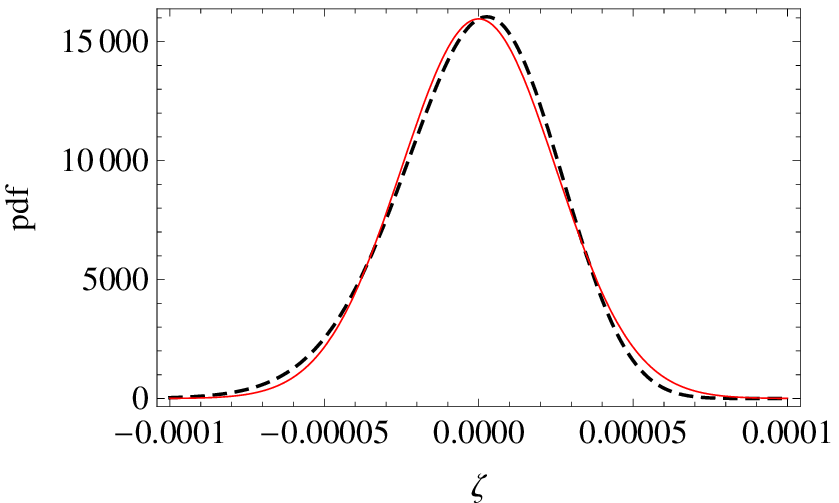}
\qquad
\includegraphics[width=0.4\textwidth]{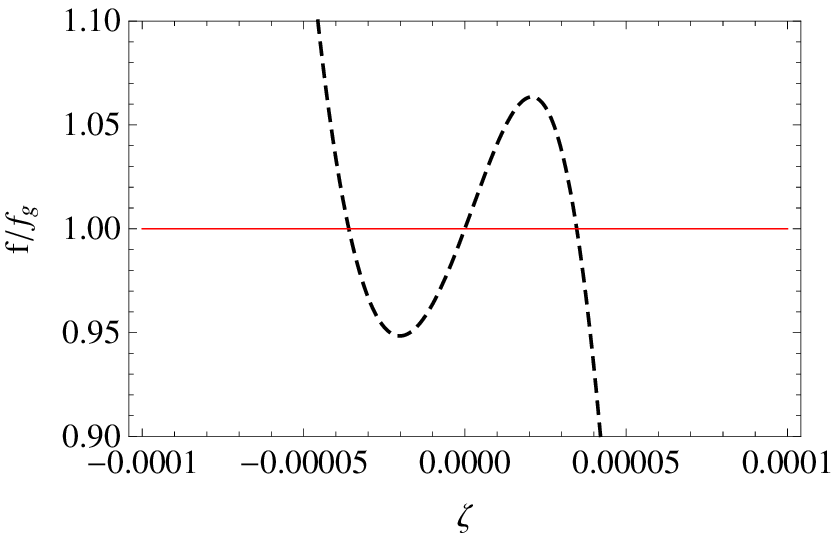}  
\qquad
\includegraphics[width=0.4\textwidth]{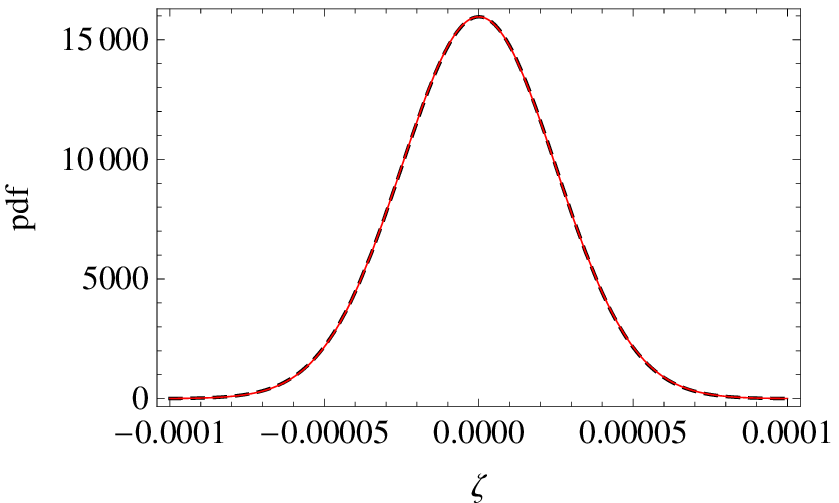}  
\qquad
\includegraphics[width=0.4\textwidth]{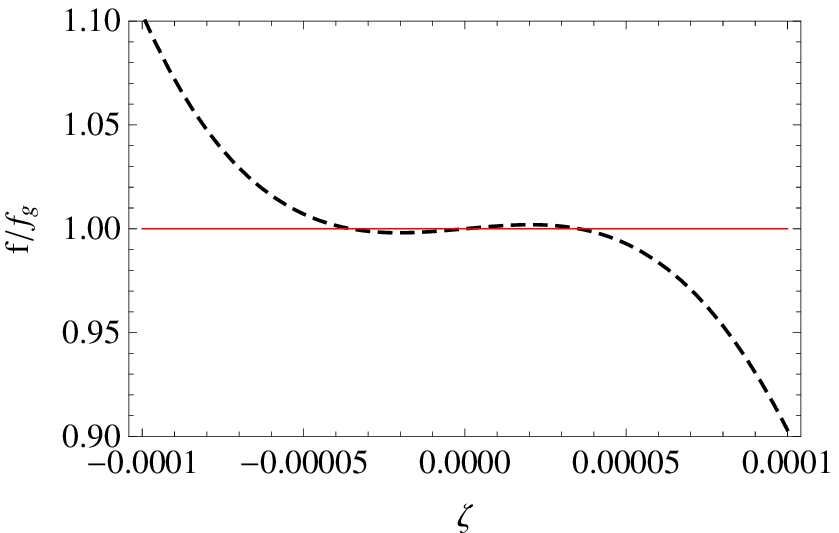}  
\caption{\label{fig::pdf3}\textit{The $D-\lambda\varphi$ model with $V\sim\cosh(m\varphi)$.} Left:  Pdfs at $f_{d}=0.8$ and $m\varphi_{*} = 0.003$ (top, $f_{NL}=3472$), $0.165$ (bottom, $f_{NL}=114$). The black dashed curve ares the pdfs $f$, while the red solid curves are the Gaussian reference, $f_{g}$. Right: The ratio of non-Gaussian pdf to the Gaussian one with $m\varphi_{*} = 0.003$ (top), $0.165$ (bottom). }
\end{center}
\end{figure}

\begin{table}[h]
\centering
  \begin{tabular}{l|l|l|l|l|l|l}
  \hline
  \hline
  \multirow{2}{*}{Moment} &  \multicolumn{3}{c|}{$m\varphi_{*}=0.03$} & \multicolumn{3}{c}{$m\varphi_{*}=0.165$}                   	\\
  \cline{2-7}
  & \multicolumn{1}{c|}{$f_{d} = 0.2$} & \multicolumn{1}{c|}{$f_{d} = 0.5$}  & \multicolumn{1}{c|}{$f_{d} = 0.8$} 
  & \multicolumn{1}{c|}{$f_{d} = 0.2$} & \multicolumn{1}{c|}{$f_{d} = 0.5$}  & \multicolumn{1}{c}{$f_{d} = 0.8$}  	\\
  \hline
  \hline
  $m(3)/\sigma^{3}$      & $ 1.4076 \times 10^{-2} $ & $ 4.3669 \times 10^{-3} $ & $ 2.6789 \times 10^{-3} $
  				     & $ 3.5764 \times 10^{-4} $ & $ 1.4406 \times 10^{-4} $ & $ 9.0822 \times 10^{-5} $   	\\
  \hline
  $m(4)/\sigma^{4}-3$   & $-2.9931$                         &$-2.9977$                          & $-2.9982$            
  				     & $-2.9985$                         &$-2.9985$                          & $-2.9985$                               	\\
  \hline
  $m(5)/\sigma^{5}$      & $ 3.4784 \times 10^{-3}$  & $ 7.4967 \times 10^{-4} $ & $ 4.3509 \times 10^{-4}$              
  				     & $ 5.5869 \times 10^{-5}$  & $ 2.2490 \times 10^{-5} $ & $ 1.4178 \times 10^{-5} $	   	\\
  \hline
  $m(6)/\sigma^{6}-15$ & $-14.999$                         & $-14.999$                         & $-14.999$                            
  				     & $-14.999$                         & $-14.999$                         & $-14.999$                                 	\\
  \hline
  \hline
  \end{tabular}
  \caption{\label{table:moment3} The value of  the third, fourth, fifth and sixth  moments of the pdf of the primordial curvature perturbation $\zeta$ with $m\varphi_{*} = 0.003, 0.165$ and $f_{d}=0.2, 0.5, 0.8$ in the $D-\lambda\varphi$ model with  $\delta\lambda_{*} =0$ and $V\sim\cosh(m\varphi)$. }
\end{table}
In Fig.~\ref{fig::pdf3}, we compare the fully non-linear pdf $f(\zeta)$ to the Gaussian $f_{g}(\zeta_{1})$ in the case when there is  large non-Gaussianity. In one of them, the non-linearity parameter is very large ($f_{d}=0.8$, $m\varphi_{*}=0.03$, $f_{NL}=3472$), and this kind of visual comparison reveals the non-Gaussianity, but in another one with  ($f_{d}=0.8$, $m\varphi_{*}=0.165$, $f_{NL}=114$),  it shows that $f$ is virtually indistinguishable from the Gaussian $f_{g}$. But, we also plot $f/f_{g}$ which reveals the non-Gaussianity in this model. In Tab.~\ref{table:moment3}, we give some values of the moments from the third up to the sixth one at $f_{d} = 0.2, 0.5, 0.8$ for $m\varphi_{*}=0.03$ and $m\varphi_{*}=0.165$ .

\section{Conclusion}
We have used $\delta \mathcal{N}$-formalism to calculate the primordial curvature perturbation for the curvaton model with a Lagrange multiplier field in two interesting cases.  We have calculate the non-linearity parameters $f_{NL}$ and $g_{NL}$ in  the probability density function of the primordial curvature perturbation  in  the sudden-decay approximation, as well as some moments of it. We find that one can get a large non-Gaussianity in this kind of model even if the curvaton dominates the total energy density before it decays, namely $f_{d}\rightarrow1$, while in the usual curvaton model with quadratic potential, one can only  get a large non-Gaussianity when the  curvaton is subdominant, namely in the limit of  $f_{d}\rightarrow 0$, see \cite{curvaton}.  It should be noticed that the isocurvature perturbations are created
when the curvaton fail to dominate the energy density while decaying.
So, it will  not produce large $f_{NL}$ comparing to that of WMAP by
taking account of  the constraint of isocurvature perturbations \cite{Mazumdar:2010sa}.  So, it seems that we have escaped the constraint
from isocurvature perturbations and give a large $f_{NL}$ comparable to
the result of WMAP. So, the introduction of Lagrange multiplier field will make the the curvaton model much richer, e.g. it also release the form of the potential. 

Furthermore, it hase been shown that with the help of a Lagrange multiplier field,  one can proposed a way to unify dark matter and dark energy in a single degree of freedom, see \cite{lagrange}. And the Lagrange multiplier modified gravity may lead to cyclic behavior very easily in the cyclic cosmology, and the scenario is much more realistic in the case of scalar cosmology \cite{Cai}.  So, we conclude that the Lagrange multiplier field could play a very interesting and important role in the construction of cosmological models, and it also interesting to study the primordial nonlinear structures and black holes  \cite{khlopov} in such kind of curvaton scenario.

\acknowledgments
We would like to thank   Anupam Mazumdar and Maxim.Yu.Khlopov for helpful  comments and discussions.
This work is supported by National Education Foundation of China grant No. 2009312711004 and Shanghai Natural Science
Foundation, China grant No. 10ZR1422000.

\appendix
\section{}


\begin{thebibliography}{999}

\bibitem{inflation}
  A.~H.~Guth,
   ``The Inflationary Universe: A Possible Solution To The Horizon And Flatness
  Phys.\ Rev.\  D {\bf 23}, 347 (1981).

  A.~D.~Linde,
   ``A New Inflationary Universe Scenario: A Possible Solution Of The Horizon,
  Phys.\ Lett.\  B {\bf 108}, 389 (1982).

  A.~J.~Albrecht and P.~J.~Steinhardt,
   ``Cosmology For Grand Unified Theories With Radiatively Induced Symmetry
  Phys.\ Rev.\ Lett.\  {\bf 48}, 1220 (1982).
  
  \bibitem{fengli}
  C.~J.~Feng, X.~Z.~Li and E.~N.~Saridakis,
  Phys.\ Rev.\  D {\bf 82}, 023526 (2010)
  [arXiv:1004.1874 [astro-ph.CO]].
  
  C.~J.~Feng and X.~Z.~Li,
  arXiv:0911.3994 [astro-ph.CO].
  
 \bibitem{Maldacena:2002vr}
  J.~M.~Maldacena,
  JHEP {\bf 0305}, 013 (2003)
  [arXiv:astro-ph/0210603].
  
  \bibitem{dwang}
  D.~Wands,
  Lect.\ Notes Phys.\  {\bf 738}, 275 (2008)
  [arXiv:astro-ph/0702187].

  
\bibitem{curvaton}
  A.~D.~Linde and V.~F.~Mukhanov,
  Phys.\ Rev.\  D {\bf 56}, 535 (1997)
  [arXiv:astro-ph/9610219].
  
  K.~Enqvist and M.~S.~Sloth,
  Nucl.\ Phys.\  B {\bf 626}, 395 (2002)
  [arXiv:hep-ph/0109214].
  
  D.~H.~Lyth and D.~Wands,
  Phys.\ Lett.\  B {\bf 524}, 5 (2002)
  [arXiv:hep-ph/0110002].
  
  T.~Moroi and T.~Takahashi,
  Phys.\ Lett.\  B {\bf 522}, 215 (2001)
  [Erratum-ibid.\  B {\bf 539}, 303 (2002)]
  [arXiv:hep-ph/0110096].
  
    \bibitem{qghuang}
  P.~Chingangbam and Q.~G.~Huang,
  arXiv:1006.4006 [astro-ph.CO].

  P.~Chingangbam and Q.~G.~Huang,
  JCAP {\bf 0904}, 031 (2009)
  [arXiv:0902.2619 [astro-ph.CO]].

  Q.~G.~Huang and Y.~Wang,
  JCAP {\bf 0809}, 025 (2008)
  [arXiv:0808.1168 [hep-th]].

  Q.~G.~Huang,
  Phys.\ Rev.\  D {\bf 78}, 043515 (2008)
  [arXiv:0807.0050 [hep-th]].

  Q.~G.~Huang,
  Phys.\ Lett.\  B {\bf 669}, 260 (2008)
  [arXiv:0801.0467 [hep-th]].

  \bibitem{piao}
  Y.~F.~Cai and Y.~Wang,
  arXiv:1005.0127 [hep-th].

  J.~Zhang, Y.~F.~Cai and Y.~S.~Piao,
  JCAP {\bf 1005}, 001 (2010)
  [arXiv:0912.0791 [hep-th]].
  
  S.~Li, Y.~F.~Cai and Y.~S.~Piao,
  Phys.\ Lett.\  B {\bf 671}, 423 (2009)
  [arXiv:0806.2363 [hep-ph]].
  
  C.~Lin and Y.~Wang,
  JCAP {\bf 1007}, 011 (2010)
  [arXiv:1004.0461 [astro-ph.CO]].

  J.~O.~Gong, C.~Lin and Y.~Wang,
  JCAP {\bf 1003}, 004 (2010)
  [arXiv:0912.2796 [astro-ph.CO]].
  
  \bibitem{other}
  C.~T.~Byrnes, K.~Enqvist and T.~Takahashi,
  arXiv:1007.5148 [astro-ph.CO].

  K.~Enqvist, A.~Mazumdar and O.~Taanila,
  arXiv:1007.0657 [astro-ph.CO].
  
  L.~Alabidi, K.~A.~Malik, C.~T.~Byrnes and K.~Y.~Choi,
  arXiv:1002.1700 [astro-ph.CO].

  A.~Mazumdar and J.~Rocher,
  arXiv:1001.0993 [hep-ph].

  S.~del Campo, R.~Herrera, J.~Saavedra, C.~Campuzano and E.~Rojas,
  Phys.\ Rev.\  D {\bf 80}, 123531 (2009)
  [arXiv:0912.4721 [astro-ph.CO]].

  K.~Enqvist, S.~Nurmi, O.~Taanila and T.~Takahashi,
  JCAP {\bf 1004}, 009 (2010)
  [arXiv:0912.4657 [astro-ph.CO]].

  J.~Sainio and I.~Vilja,
  Phys.\ Rev.\  D {\bf 81}, 083516 (2010)
  [arXiv:0912.3394 [astro-ph.CO]].
  
  C.~M.~Lin and K.~Cheung,
  arXiv:0911.4749 [hep-ph].

  K.~Nakayama and J.~Yokoyama,
  JCAP {\bf 1001}, 010 (2010)
  [arXiv:0910.0715 [astro-ph.CO]].

  K.~Enqvist and T.~Takahashi,
  JCAP {\bf 0912}, 001 (2009)
  [arXiv:0909.5362 [astro-ph.CO]].

  A.~Chambers, S.~Nurmi and A.~Rajantie,
  JCAP {\bf 1001}, 012 (2010)
  [arXiv:0909.4535 [astro-ph.CO]].

  K.~Dimopoulos, M.~Karciauskas and J.~M.~Wagstaff,
  Phys.\ Lett.\  B {\bf 683}, 298 (2010)
  [arXiv:0909.0475 [hep-ph]].

  K.~Dimopoulos, M.~Karciauskas and J.~M.~Wagstaff,
  Phys.\ Rev.\  D {\bf 81}, 023522 (2010)
  [arXiv:0907.1838 [hep-ph]].

  K.~Enqvist, S.~Nurmi, G.~Rigopoulos, O.~Taanila and T.~Takahashi,
  JCAP {\bf 0911}, 003 (2009)
  [arXiv:0906.3126 [astro-ph.CO]].

  C.~M.~Lin and K.~Cheung,
  JCAP {\bf 0906}, 006 (2009)
  [arXiv:0904.2826 [hep-ph]].
  
  \bibitem{WMAP3}
  D.~N.~Spergel {\it et al.}  [WMAP Collaboration],
  Astrophys.\ J.\ Suppl.\  {\bf 170}, 377 (2007)
  [arXiv:astro-ph/0603449].
  
  \bibitem{WMAP5}
  E.~Komatsu {\it et al.}  [WMAP Collaboration],
  Astrophys.\ J.\ Suppl.\  {\bf 180}, 330 (2009)
  [arXiv:0803.0547 [astro-ph]].
  
  \bibitem{WMAP7}
  E.~Komatsu {\it et al.},
  arXiv:1001.4538 [astro-ph.CO].
  
  \bibitem{lagrange}
  C.~Gao, Y.~Gong, X.~Wang and X.~Chen,
  arXiv:1003.6056 [astro-ph.CO].
  
  E.~A.~Lim, I.~Sawicki and A.~Vikman,
  JCAP {\bf 1005}, 012 (2010)
  [arXiv:1003.5751 [astro-ph.CO]].
  
  \bibitem{deltan}
  A.~A.~Starobinsky,
   ``Multicomponent de Sitter (Inflationary) Stages and the Generation of
  JETP Lett.\  {\bf 42}, 152 (1985)
  [Pisma Zh.\ Eksp.\ Teor.\ Fiz.\  {\bf 42}, 124 (1985)].

  M.~Sasaki and E.~D.~Stewart,
   ``A General Analytic Formula For The Spectral Index Of The Density
  Prog.\ Theor.\ Phys.\  {\bf 95}, 71 (1996)
  [arXiv:astro-ph/9507001].

  D.~H.~Lyth and Y.~Rodriguez,
  Phys.\ Rev.\ Lett.\  {\bf 95}, 121302 (2005)
  [arXiv:astro-ph/0504045].

  \bibitem{Lyth:2004gb}
  D.~H.~Lyth, K.~A.~Malik and M.~Sasaki,
  JCAP {\bf 0505}, 004 (2005)
  [arXiv:astro-ph/0411220].
  
  \bibitem{Sasaki:2006kq}
  M.~Sasaki, J.~Valiviita and D.~Wands,
  Phys.\ Rev.\  D {\bf 74}, 103003 (2006)
  [arXiv:astro-ph/0607627].
 
  \bibitem{Huang:2008zj}
  Q.~G.~Huang,
  JCAP {\bf 0811}, 005 (2008)
  [arXiv:0808.1793 [hep-th]].
  
  \bibitem{Mazumdar:2010sa}
  A.~Mazumdar and J.~Rocher,
  arXiv:1001.0993 [hep-ph].
  
  \bibitem{Cai}
  Y.~F.~Cai and E.~N.~Saridakis,
  arXiv:1007.3204 [astro-ph.CO].
  
  \bibitem{khlopov}
  A.S.Sakharov and M.Yu.Khlopov, 
 Yadernaya Fizika (1994) V. 57, PP. 514- 516. [English translation: Phys.Atom.Nucl. (1994) V.
57, PP. 485-487];

A.S.Sakharov, D.D.Sokoloff and M.Yu.Khlopov
Yadernaya Fizika (1996) V. 59, PP.
1050-1055. [English translation: Phys.Atom.Nucl. (1996) V. 59, PP. 1005-1010];

   M.Yu.Khlopov, A.S.Sakharov and D.D.Sokoloff
 Nucl.Phys. B (Proc. Suppl.) (1999) V. 72, 105-109;

 Sergei G. Rubin, Alexander, S.Sakharov, Maxim Yu. Khlopov, J.Exp.Theor.Phys.91:921-929,2001,
[hep-ph/0106187];

 M.Yu. Khlopov, S.G. Rubin, A.S. Sakharov, CERN-TH-2002-033, Feb 2002.14pp.
 Grav. \& Cosmol., v.8, Suppl. 2002, pp.57-65 [astro-ph/0202505];

 Maxim.Yu. Khlopov, Sergei.G. Rubin, Alexander.S. Sakharov
 Astropart. Phys. (2005) V. 23, N-2, PP. 265-277. [astro-ph/0401532];

 M.Yu.Khlopov, S.G.Rubin, Kluwer
 Academic Publishers, Dordrecht, 295 pp., 2004;
 
 M.Yu.Khlopov, Conference series. (2007) V.66, P.012032 (10 pages). XXIX
Spanish relativity meeting (ERE2006);

 M.Yu.Khlopov, Res.Astron.Astrophys. (2010) V. 10, PP. 495-528, [arXiv:0801.0116].
  


\end{thebibliography}
\end{document}